\pdfoutput=1
\documentclass[11pt,twoside,a4paper,pdftex,cmspaper,final,collab]{cms-tdr}

\begin{document}\cmsNoteHeader{CFT-09-003}
%
%
%

%
%
\hyphenation{env-iron-men-tal}
\hyphenation{had-ron-i-za-tion}
\hyphenation{cal-or-i-me-ter}
\hyphenation{de-vices}
%
%
\RCS$Revision: 1.26 $
\RCS$Date: 2009/11/13 21:26:53 $
\RCS$Name:  $
\newcommand{\gevc}{\ensuremath{\mathrm{GeV}/c}}
\newcommand{\celsius}{\ensuremath{^\circ C}}
\newcommand{\degree}{\ensuremath{^\circ}}

%
%
%

\newcommand {\etal}{\mbox{et al.}\xspace} 
\newcommand {\ie}{\mbox{i.e.}\xspace}     
\newcommand {\eg}{\mbox{e.g.}\xspace}     
\newcommand {\etc}{\mbox{etc.}\xspace}     
\newcommand {\vs}{\mbox{\sl vs.}\xspace}      
\newcommand {\mdash}{\ensuremath{\mathrm{-}}} 

\newcommand {\Lone}{Level-1\xspace} 
\newcommand {\Ltwo}{Level-2\xspace}
\newcommand {\Lthree}{Level-3\xspace}

\providecommand{\ACERMC} {\textsc{AcerMC}\xspace}
\providecommand{\ALPGEN} {{\textsc{alpgen}}\xspace}
\providecommand{\CHARYBDIS} {{\textsc{charybdis}}\xspace}
\providecommand{\CMKIN} {\textsc{cmkin}\xspace}
\providecommand{\CMSIM} {{\textsc{cmsim}}\xspace}
\providecommand{\CMSSW} {{\textsc{cmssw}}\xspace}
\providecommand{\COBRA} {{\textsc{cobra}}\xspace}
\providecommand{\COCOA} {{\textsc{cocoa}}\xspace}
\providecommand{\COMPHEP} {\textsc{CompHEP}\xspace}
\providecommand{\EVTGEN} {{\textsc{evtgen}}\xspace}
\providecommand{\FAMOS} {{\textsc{famos}}\xspace}
\providecommand{\GARCON} {\textsc{garcon}\xspace}
\providecommand{\GARFIELD} {{\textsc{garfield}}\xspace}
\providecommand{\GEANE} {{\textsc{geane}}\xspace}
\providecommand{\GEANTfour} {{\textsc{geant4}}\xspace}
\providecommand{\GEANTthree} {{\textsc{geant3}}\xspace}
\providecommand{\GEANT} {{\textsc{geant}}\xspace}
\providecommand{\HDECAY} {\textsc{hdecay}\xspace}
\providecommand{\HERWIG} {{\textsc{herwig}}\xspace}
\providecommand{\HIGLU} {{\textsc{higlu}}\xspace}
\providecommand{\HIJING} {{\textsc{hijing}}\xspace}
\providecommand{\IGUANA} {\textsc{iguana}\xspace}
\providecommand{\ISAJET} {{\textsc{isajet}}\xspace}
\providecommand{\ISAPYTHIA} {{\textsc{isapythia}}\xspace}
\providecommand{\ISASUGRA} {{\textsc{isasugra}}\xspace}
\providecommand{\ISASUSY} {{\textsc{isasusy}}\xspace}
\providecommand{\ISAWIG} {{\textsc{isawig}}\xspace}
\providecommand{\MADGRAPH} {\textsc{MadGraph}\xspace}
\providecommand{\MCATNLO} {\textsc{mc@nlo}\xspace}
\providecommand{\MCFM} {\textsc{mcfm}\xspace}
\providecommand{\MILLEPEDE} {{\textsc{millepede}}\xspace}
\providecommand{\ORCA} {{\textsc{orca}}\xspace}
\providecommand{\OSCAR} {{\textsc{oscar}}\xspace}
\providecommand{\PHOTOS} {\textsc{photos}\xspace}
\providecommand{\PROSPINO} {\textsc{prospino}\xspace}
\providecommand{\PYTHIA} {{\textsc{pythia}}\xspace}
\providecommand{\SHERPA} {{\textsc{sherpa}}\xspace}
\providecommand{\TAUOLA} {\textsc{tauola}\xspace}
\providecommand{\TOPREX} {\textsc{TopReX}\xspace}
\providecommand{\XDAQ} {{\textsc{xdaq}}\xspace}

\newcommand {\DZERO}{D\O\xspace}     


\newcommand{\de}{\ensuremath{^\circ}}
\newcommand{\ten}[1]{\ensuremath{\times \text{10}^\text{#1}}}
\newcommand{\unit}[1]{\ensuremath{\text{\,#1}}\xspace}
\newcommand{\mum}{\ensuremath{\,\mu\text{m}}\xspace}
\newcommand{\micron}{\ensuremath{\,\mu\text{m}}\xspace}
\newcommand{\cm}{\ensuremath{\,\text{cm}}\xspace}
\newcommand{\mm}{\ensuremath{\,\text{mm}}\xspace}
\newcommand{\mus}{\ensuremath{\,\mu\text{s}}\xspace}
\newcommand{\keV}{\ensuremath{\,\text{ke\hspace{-.08em}V}}\xspace}
\newcommand{\MeV}{\ensuremath{\,\text{Me\hspace{-.08em}V}}\xspace}
\newcommand{\GeV}{\ensuremath{\,\text{Ge\hspace{-.08em}V}}\xspace}
\newcommand{\TeV}{\ensuremath{\,\text{Te\hspace{-.08em}V}}\xspace}
\newcommand{\PeV}{\ensuremath{\,\text{Pe\hspace{-.08em}V}}\xspace}
\newcommand{\keVc}{\ensuremath{{\,\text{ke\hspace{-.08em}V\hspace{-0.16em}/\hspace{-0.08em}c}}}\xspace}
\newcommand{\MeVc}{\ensuremath{{\,\text{Me\hspace{-.08em}V\hspace{-0.16em}/\hspace{-0.08em}c}}}\xspace}
\newcommand{\GeVc}{\ensuremath{{\,\text{Ge\hspace{-.08em}V\hspace{-0.16em}/\hspace{-0.08em}c}}}\xspace}
\newcommand{\TeVc}{\ensuremath{{\,\text{Te\hspace{-.08em}V\hspace{-0.16em}/\hspace{-0.08em}c}}}\xspace}
\newcommand{\keVcc}{\ensuremath{{\,\text{ke\hspace{-.08em}V\hspace{-0.16em}/\hspace{-0.08em}c}^\text{2}}}\xspace}
\newcommand{\MeVcc}{\ensuremath{{\,\text{Me\hspace{-.08em}V\hspace{-0.16em}/\hspace{-0.08em}c}^\text{2}}}\xspace}
\newcommand{\GeVcc}{\ensuremath{{\,\text{Ge\hspace{-.08em}V\hspace{-0.16em}/\hspace{-0.08em}c}^\text{2}}}\xspace}
\newcommand{\TeVcc}{\ensuremath{{\,\text{Te\hspace{-.08em}V\hspace{-0.16em}/\hspace{-0.08em}c}^\text{2}}}\xspace}

\newcommand{\pbinv} {\mbox{\ensuremath{\,\text{pb}^\text{$-$1}}}\xspace}
\newcommand{\fbinv} {\mbox{\ensuremath{\,\text{fb}^\text{$-$1}}}\xspace}
\newcommand{\nbinv} {\mbox{\ensuremath{\,\text{nb}^\text{$-$1}}}\xspace}
\newcommand{\percms}{\ensuremath{\,\text{cm}^\text{$-$2}\,\text{s}^\text{$-$1}}\xspace}
\newcommand{\lumi}{\ensuremath{\mathcal{L}}\xspace}
\newcommand{\Lumi}{\ensuremath{\mathcal{L}}\xspace}
%
\newcommand{\LvLow}  {\ensuremath{\mathcal{L}=\text{10}^\text{32}\,\text{cm}^\text{$-$2}\,\text{s}^\text{$-$1}}\xspace}
\newcommand{\LLow}   {\ensuremath{\mathcal{L}=\text{10}^\text{33}\,\text{cm}^\text{$-$2}\,\text{s}^\text{$-$1}}\xspace}
\newcommand{\lowlumi}{\ensuremath{\mathcal{L}=\text{2}\times \text{10}^\text{33}\,\text{cm}^\text{$-$2}\,\text{s}^\text{$-$1}}\xspace}
\newcommand{\LMed}   {\ensuremath{\mathcal{L}=\text{2}\times \text{10}^\text{33}\,\text{cm}^\text{$-$2}\,\text{s}^\text{$-$1}}\xspace}
\newcommand{\LHigh}  {\ensuremath{\mathcal{L}=\text{10}^\text{34}\,\text{cm}^\text{$-$2}\,\text{s}^\text{$-$1}}\xspace}
\newcommand{\hilumi} {\ensuremath{\mathcal{L}=\text{10}^\text{34}\,\text{cm}^\text{$-$2}\,\text{s}^\text{$-$1}}\xspace}


\newcommand{\zp}{\ensuremath{\mathrm{Z}^\prime}\xspace}


\newcommand{\kt}{\ensuremath{k_{\mathrm{T}}}\xspace}
\newcommand{\BC}{\ensuremath{{B_{\mathrm{c}}}}\xspace}
\newcommand{\bbarc}{\ensuremath{{\overline{b}c}}\xspace}
\newcommand{\bbbar}{\ensuremath{{b\overline{b}}}\xspace}
\newcommand{\ccbar}{\ensuremath{{c\overline{c}}}\xspace}
\newcommand{\JPsi}{\ensuremath{{J}/\psi}\xspace}
\newcommand{\bspsiphi}{\ensuremath{B_s \to \JPsi\, \phi}\xspace}
\newcommand{\AFB}{\ensuremath{A_\mathrm{FB}}\xspace}
\newcommand{\EE}{\ensuremath{e^+e^-}\xspace}
\newcommand{\MM}{\ensuremath{\mu^+\mu^-}\xspace}
\newcommand{\TT}{\ensuremath{\tau^+\tau^-}\xspace}
\newcommand{\wangle}{\ensuremath{\sin^{2}\theta_{\mathrm{eff}}^\mathrm{lept}(M^2_\mathrm{Z})}\xspace}
\newcommand{\ttbar}{\ensuremath{{t\overline{t}}}\xspace}
\newcommand{\stat}{\ensuremath{\,\text{(stat.)}}\xspace}
\newcommand{\syst}{\ensuremath{\,\text{(syst.)}}\xspace}

\newcommand{\HGG}{\ensuremath{\mathrm{H}\to\gamma\gamma}}
\newcommand{\gev}{\GeV}
\newcommand{\GAMJET}{\ensuremath{\gamma + \mathrm{jet}}}
\newcommand{\PPTOJETS}{\ensuremath{\mathrm{pp}\to\mathrm{jets}}}
\newcommand{\PPTOGG}{\ensuremath{\mathrm{pp}\to\gamma\gamma}}
\newcommand{\PPTOGAMJET}{\ensuremath{\mathrm{pp}\to\gamma +
\mathrm{jet}
}}
\newcommand{\MH}{\ensuremath{\mathrm{M_{\mathrm{H}}}}}
\newcommand{\RNINE}{\ensuremath{\mathrm{R}_\mathrm{9}}}
\newcommand{\DR}{\ensuremath{\Delta\mathrm{R}}}


\newcommand{\PT}{\ensuremath{p_{\mathrm{T}}}\xspace}
\newcommand{\pt}{\ensuremath{p_{\mathrm{T}}}\xspace}
\newcommand{\ET}{\ensuremath{E_{\mathrm{T}}}\xspace}
\newcommand{\HT}{\ensuremath{H_{\mathrm{T}}}\xspace}
\newcommand{\et}{\ensuremath{E_{\mathrm{T}}}\xspace}
\newcommand{\Em}{\ensuremath{E\!\!\!/}\xspace}
\newcommand{\Pm}{\ensuremath{p\!\!\!/}\xspace}
\newcommand{\PTm}{\ensuremath{{p\!\!\!/}_{\mathrm{T}}}\xspace}
\newcommand{\ETm}{\ensuremath{E_{\mathrm{T}}^{\mathrm{miss}}}\xspace}
\newcommand{\MET}{\ensuremath{E_{\mathrm{T}}^{\mathrm{miss}}}\xspace}
\newcommand{\ETmiss}{\ensuremath{E_{\mathrm{T}}^{\mathrm{miss}}}\xspace}
\newcommand{\VEtmiss}{\ensuremath{{\vec E}_{\mathrm{T}}^{\mathrm{miss}}}\xspace}

%

\newcommand{\ga}{\ensuremath{\gtrsim}}
\newcommand{\la}{\ensuremath{\lesssim}}
\newcommand{\swsq}{\ensuremath{\sin^2\theta_W}\xspace}
\newcommand{\cwsq}{\ensuremath{\cos^2\theta_W}\xspace}
\newcommand{\tanb}{\ensuremath{\tan\beta}\xspace}
\newcommand{\tanbsq}{\ensuremath{\tan^{2}\beta}\xspace}
\newcommand{\sidb}{\ensuremath{\sin 2\beta}\xspace}
\newcommand{\alpS}{\ensuremath{\alpha_S}\xspace}
\newcommand{\alpt}{\ensuremath{\tilde{\alpha}}\xspace}

\newcommand{\QL}{\ensuremath{Q_L}\xspace}
\newcommand{\sQ}{\ensuremath{\tilde{Q}}\xspace}
\newcommand{\sQL}{\ensuremath{\tilde{Q}_L}\xspace}
\newcommand{\ULC}{\ensuremath{U_L^C}\xspace}
\newcommand{\sUC}{\ensuremath{\tilde{U}^C}\xspace}
\newcommand{\sULC}{\ensuremath{\tilde{U}_L^C}\xspace}
\newcommand{\DLC}{\ensuremath{D_L^C}\xspace}
\newcommand{\sDC}{\ensuremath{\tilde{D}^C}\xspace}
\newcommand{\sDLC}{\ensuremath{\tilde{D}_L^C}\xspace}
\newcommand{\LL}{\ensuremath{L_L}\xspace}
\newcommand{\sL}{\ensuremath{\tilde{L}}\xspace}
\newcommand{\sLL}{\ensuremath{\tilde{L}_L}\xspace}
\newcommand{\ELC}{\ensuremath{E_L^C}\xspace}
\newcommand{\sEC}{\ensuremath{\tilde{E}^C}\xspace}
\newcommand{\sELC}{\ensuremath{\tilde{E}_L^C}\xspace}
\newcommand{\sEL}{\ensuremath{\tilde{E}_L}\xspace}
\newcommand{\sER}{\ensuremath{\tilde{E}_R}\xspace}
\newcommand{\sFer}{\ensuremath{\tilde{f}}\xspace}
\newcommand{\sQua}{\ensuremath{\tilde{q}}\xspace}
\newcommand{\sUp}{\ensuremath{\tilde{u}}\xspace}
\newcommand{\suL}{\ensuremath{\tilde{u}_L}\xspace}
\newcommand{\suR}{\ensuremath{\tilde{u}_R}\xspace}
\newcommand{\sDw}{\ensuremath{\tilde{d}}\xspace}
\newcommand{\sdL}{\ensuremath{\tilde{d}_L}\xspace}
\newcommand{\sdR}{\ensuremath{\tilde{d}_R}\xspace}
\newcommand{\sTop}{\ensuremath{\tilde{t}}\xspace}
\newcommand{\stL}{\ensuremath{\tilde{t}_L}\xspace}
\newcommand{\stR}{\ensuremath{\tilde{t}_R}\xspace}
\newcommand{\stone}{\ensuremath{\tilde{t}_1}\xspace}
\newcommand{\sttwo}{\ensuremath{\tilde{t}_2}\xspace}
\newcommand{\sBot}{\ensuremath{\tilde{b}}\xspace}
\newcommand{\sbL}{\ensuremath{\tilde{b}_L}\xspace}
\newcommand{\sbR}{\ensuremath{\tilde{b}_R}\xspace}
\newcommand{\sbone}{\ensuremath{\tilde{b}_1}\xspace}
\newcommand{\sbtwo}{\ensuremath{\tilde{b}_2}\xspace}
\newcommand{\sLep}{\ensuremath{\tilde{l}}\xspace}
\newcommand{\sLepC}{\ensuremath{\tilde{l}^C}\xspace}
\newcommand{\sEl}{\ensuremath{\tilde{e}}\xspace}
\newcommand{\sElC}{\ensuremath{\tilde{e}^C}\xspace}
\newcommand{\seL}{\ensuremath{\tilde{e}_L}\xspace}
\newcommand{\seR}{\ensuremath{\tilde{e}_R}\xspace}
\newcommand{\snL}{\ensuremath{\tilde{\nu}_L}\xspace}
\newcommand{\sMu}{\ensuremath{\tilde{\mu}}\xspace}
\newcommand{\sNu}{\ensuremath{\tilde{\nu}}\xspace}
\newcommand{\sTau}{\ensuremath{\tilde{\tau}}\xspace}
\newcommand{\Glu}{\ensuremath{g}\xspace}
\newcommand{\sGlu}{\ensuremath{\tilde{g}}\xspace}
\newcommand{\Wpm}{\ensuremath{W^{\pm}}\xspace}
\newcommand{\sWpm}{\ensuremath{\tilde{W}^{\pm}}\xspace}
\newcommand{\Wz}{\ensuremath{W^{0}}\xspace}
\newcommand{\sWz}{\ensuremath{\tilde{W}^{0}}\xspace}
\newcommand{\sWino}{\ensuremath{\tilde{W}}\xspace}
\newcommand{\Bz}{\ensuremath{B^{0}}\xspace}
\newcommand{\sBz}{\ensuremath{\tilde{B}^{0}}\xspace}
\newcommand{\sBino}{\ensuremath{\tilde{B}}\xspace}
\newcommand{\Zz}{\ensuremath{Z^{0}}\xspace}
\newcommand{\sZino}{\ensuremath{\tilde{Z}^{0}}\xspace}
\newcommand{\sGam}{\ensuremath{\tilde{\gamma}}\xspace}
\newcommand{\chiz}{\ensuremath{\tilde{\chi}^{0}}\xspace}
\newcommand{\chip}{\ensuremath{\tilde{\chi}^{+}}\xspace}
\newcommand{\chim}{\ensuremath{\tilde{\chi}^{-}}\xspace}
\newcommand{\chipm}{\ensuremath{\tilde{\chi}^{\pm}}\xspace}
\newcommand{\Hone}{\ensuremath{H_{d}}\xspace}
\newcommand{\sHone}{\ensuremath{\tilde{H}_{d}}\xspace}
\newcommand{\Htwo}{\ensuremath{H_{u}}\xspace}
\newcommand{\sHtwo}{\ensuremath{\tilde{H}_{u}}\xspace}
\newcommand{\sHig}{\ensuremath{\tilde{H}}\xspace}
\newcommand{\sHa}{\ensuremath{\tilde{H}_{a}}\xspace}
\newcommand{\sHb}{\ensuremath{\tilde{H}_{b}}\xspace}
\newcommand{\sHpm}{\ensuremath{\tilde{H}^{\pm}}\xspace}
\newcommand{\hz}{\ensuremath{h^{0}}\xspace}
\newcommand{\Hz}{\ensuremath{H^{0}}\xspace}
\newcommand{\Az}{\ensuremath{A^{0}}\xspace}
\newcommand{\Hpm}{\ensuremath{H^{\pm}}\xspace}
\newcommand{\sGra}{\ensuremath{\tilde{G}}\xspace}
\newcommand{\mtil}{\ensuremath{\tilde{m}}\xspace}
\newcommand{\rpv}{\ensuremath{\rlap{\kern.2em/}R}\xspace}
\newcommand{\LLE}{\ensuremath{LL\bar{E}}\xspace}
\newcommand{\LQD}{\ensuremath{LQ\bar{D}}\xspace}
\newcommand{\UDD}{\ensuremath{\overline{UDD}}\xspace}
\newcommand{\Lam}{\ensuremath{\lambda}\xspace}
\newcommand{\Lamp}{\ensuremath{\lambda'}\xspace}
\newcommand{\Lampp}{\ensuremath{\lambda''}\xspace}
\newcommand{\spinbd}[2]{\ensuremath{\bar{#1}_{\dot{#2}}}\xspace}

\newcommand{\MD}{\ensuremath{{M_\mathrm{D}}}\xspace}
\newcommand{\Mpl}{\ensuremath{{M_\mathrm{Pl}}}\xspace}
\newcommand{\Rinv} {\ensuremath{{R}^{-1}}\xspace}

%
%
\hyphenation{en-viron-men-tal}

\cmsNoteHeader{09-003}
\title{
Alignment of the CMS Silicon Tracker during Commissioning with Cosmic Rays}

\author{CMS collaboration}

\date{\today}

\abstract{
The CMS silicon tracker, consisting of 1440 silicon pixel and 
15\,148 silicon strip detector modules, 
has been aligned using more than three million cosmic ray charged  
particles, with additional information from optical surveys.
The positions of the modules were determined  with respect  to cosmic ray 
trajectories to an average precision of 3--4 microns RMS in the barrel 
and 3--14 microns RMS in the endcap in the most sensitive  coordinate. 
The results have been validated by several studies, including laser beam 
cross-checks, track fit self-consistency, track residuals in overlapping module 
regions, and track parameter resolution, and are compared with predictions 
obtained from simulation. Correlated systematic effects have been investigated. 
The track  parameter resolutions obtained with this alignment are close 
to the design performance.
}

\hypersetup{%
pdfauthor={CMS Collaboration},%
pdftitle={Alignment of the CMS Silicon Tracker during Commissioning with Cosmic Rays},%
pdfsubject={CMS},%
pdfkeywords={CMS, CRAFT, tracker, alignment, silicon strip detectors, silicon pixel detectors, tracking systems}}

\maketitle 

\section{Introduction}
\label{sec:intro}

The main goal of the CMS experiment~\cite{detector-paper} is to
explore physics at the TeV energy scale exploiting the proton-proton
collisions delivered by the Large Hadron Collider (LHC)~\cite{lhc-paper}.
The CMS silicon tracking detector (tracker)~\cite{trackertdr,trackertdraddendum} consists of 
1440 silicon pixel and 15\,148 silicon strip detector modules and is located,  
together with the electromagnetic and hadronic calorimeters, inside a
superconducting solenoidal magnet operating at a field of
3.8\,T. Outside of the solenoid, the muon system is used both for triggering 
on muons and reconstructing their trajectories.

The aim of the tracker is to measure the trajectories of charged particles (tracks)
with excellent momentum, angle, and position resolution, and with high
pattern recognition efficiency~\cite{cmstdr}.   
Precise determination of the position of all silicon modules is a challenging task and
one of the critical aspects for achieving the design track parameter
resolutions. In the context of alignment, {\it position} is used throughout
this paper to refer both to the location of the center point of active areas
on modules and to the orientation of active areas.  The hit position
resolution is in the range 10 to 30~$\mu$m and therefore the alignment
precision should be better than 10~$\mu$m to achieve optimal track parameter
resolution.  Simulation studies based on the design (ideal) tracker geometry
imply an impact parameter resolution of about 15~$\mu$m and a transverse
momentum ($p_T$) resolution of about 1.5\% for 100 GeV/$c$ muons.

The CMS collaboration conducted a month-long data-taking exercise
known as the Cosmic Run At Four Tesla (CRAFT) during October-November 2008
with the goal of commissioning the experiment for extended data-taking~\cite{craftgeneral}. 
With all installed detector systems participating, CMS recorded 270 million
cosmic-ray-triggered events with the solenoid at its nominal axial field
strength of 3.8~T. However, only a few percent of those events had cosmic rays 
traversing the tracker volume. Prior to CRAFT and during the final installation phase 
of the experiment, a series of commissioning exercises to record cosmic ray events 
took place with the solenoid turned off from May to September 2008. 
The tracker was included only at the later stages of those runs and a smaller
number of cosmic ray events with the tracker participation were recorded 
compared with CRAFT. The operating temperature of the tracker during the 
CRAFT data-taking period was stable at around room temperature. Excellent 
performance of the tracking system has been achieved with both the silicon
strip~\cite{crafttracking} and silicon pixel~\cite{craftpixel} components.

In this paper, the measurement of detector module positions for the
full tracker from alignment analysis with cosmic ray particles, 
optical survey information, and the Laser Alignment System (LAS) is presented. 
This work builds on the experience with the alignment of about 15\% of the CMS
silicon strip tracker during stand-alone commissioning~\cite{tifalignment}.
Significant improvements have been introduced. Inclusion of the pixel detector
and of the full strip tracker made the first alignment of the whole CMS tracker possible.
The magnetic field allowed estimates of the multiple scattering
effects on a track-by-track basis. 
The data taken with the solenoid turned off were used in this analysis to
cross-check the field-on data in a small number of cases.
Significantly more tracks in a
single stable configuration improved statistical precision. Two complementary
statistical alignment algorithms, a global and a local iterative method,  were
applied to the data. Although both algorithms had been  
used previously during stand-alone commissioning,
further improvements have been introduced, such as inclusion of
survey data together with tracks. 
Results from the alignment studies have been compared extensively with
detailed Monte Carlo (MC) simulations of the detector.

This paper is structured as follows: in Section \ref{sec:layout}, the tracker
layout and coordinate systems are introduced. Section \ref{sec:survey}
summarizes the optical survey measurements. In Section \ref{sec:stat}, the
track reconstruction and alignment algorithms are described and their
application to data is explained. Results are presented in
Section \ref{sec:valid}. 

\section{Tracker Layout and the Coordinate Systems}
\label{sec:layout}

\begin{figure}[bt] 
\begin{center}
\includegraphics*[width=0.75\textwidth]{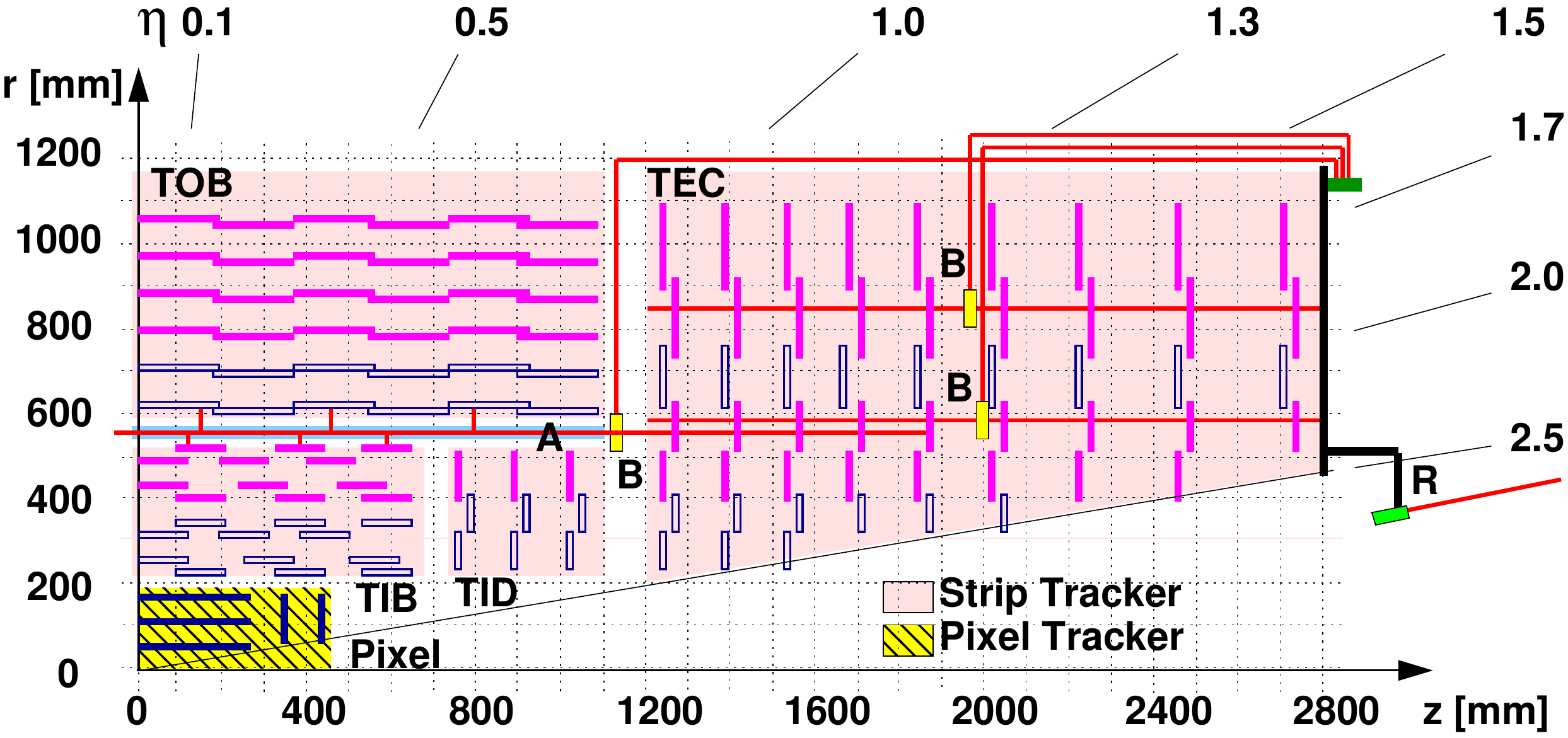}
\caption{
A quarter of the CMS silicon tracker in an $rz$ view. 
Single-sided silicon strip module positions are indicated as solid light (purple) lines, 
double-sided strip modules as open (blue) lines,
and pixel modules as solid dark (blue) lines.
Also shown are the paths of the laser rays (R), 
the beam splitters (B), and the alignment tubes (A)
of the Laser Alignment System.
\label{fig:stripLayout}
}
\end{center}
\end{figure}

CMS uses a right-handed coordinate system, with the origin at the nominal
collision point, the $x$-axis pointing to the center of the LHC, the $y$-axis
pointing up (perpendicular to the LHC plane), and the $z$-axis along the
anticlockwise-beam direction.  The polar angle ($\theta$) is measured from the
positive $z$-axis and the azimuthal angle ($\phi$) is measured from the
positive $x$-axis in the $x$-$y$ plane, whereas the radius ($r$) denotes the
distance from the $z$-axis.

The CMS tracker consists of a silicon pixel detector
and a silicon strip detector (Fig.~\ref{fig:stripLayout}).
The silicon pixel detector is composed of two sub-detectors,
the barrel (BPIX) and the two endcaps in the forward regions (FPIX). 
The pixel modules provide two-dimensional measurements of the hit position
in the module planes, which effectively translate into three-dimensional
measurements in space.
The silicon strip detector is composed of four
sub-detectors: the Tracker Inner and Outer Barrels (TIB and TOB), the Tracker
Inner Disks (TID), and the Tracker Endcaps (TEC). All sub-detectors are
concentrically arranged around the nominal beam axis. 
The two inner layers of both the TIB and
TOB, the two inner rings of the TID, and the first, second, and fifth rings of the TEC
are equipped with double-sided modules; all other positions have single-sided
modules. Single-sided modules provide $r\phi$ measurements in the
barrel and $\phi$ measurements in the endcaps.
Double-sided modules are made of a pair of single-sided
strip modules, one $r\phi$ and one stereo module in the barrel, and
one $\phi$ and one stereo module in the endcaps, precisely mounted
back-to-back, with the stereo module sensors tilted by 100\,mrad. 
For simplicity, we refer to both $r\phi$ and $\phi$ modules as $r\phi$ 
in the rest of the paper.

In the barrel region, modules are arranged in linear structures parallel to the $z$-axis,
such as ladders in the BPIX, strings in the TIB, and rods in the TOB.
The endcaps are composed of disks, which in turn contain wedge-shaped structures 
covering a narrow $\phi$ region, 
such as blades in the FPIX and petals in the TEC.
The BPIX and the TIB are composed of two half-barrel structures, 
separated along the $x=0$ plane for the BPIX and the $z=0$ plane for the TIB.

\begin{figure}[tbh]
\begin{center}
\includegraphics*[height=4cm]{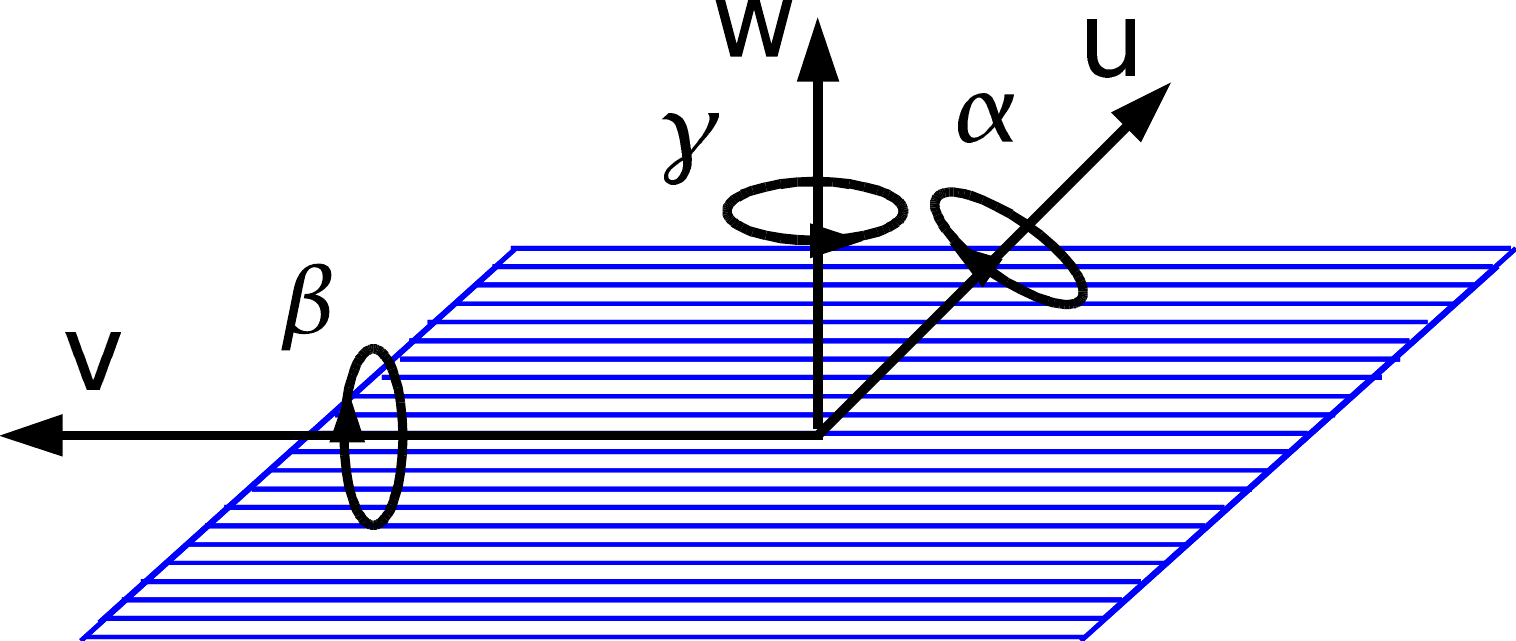}
\caption{
Illustration of the module local coordinates $u,v,w$ and the corresponding
rotations $\alpha,\beta,\gamma$ for a single-sided strip module.
}
\label{fig:coordinates}
\end{center}
\end{figure}

A local right-handed coordinate system is defined for each module with
the origin at the geometric center of the active area of the module.
As illustrated in Fig.~\ref{fig:coordinates},
the $u$-axis is defined along the more precisely
measured coordinate of the module (typically along the azimuthal direction in
the global system), the $v$-axis orthogonal to the $u$-axis and in the module
plane, pointing away from the readout electronics, and the $w$-axis normal to
the module plane. When double-sided modules are considered as a single entity,
the coordinate system is referenced to the $r\phi$ module.
For the pixel system, $u$ is always orthogonal to the magnetic field, that is
in global $r\phi$ direction in the BPIX and in the radial direction in the FPIX.
The $v$ coordinate is perpendicular to $u$ in the sensor plane, that is along
global $z$ in the BPIX and at a small angle to the global $r\phi$ direction in
the FPIX.  The angles $\alpha$, $\beta$, and $\gamma$ indicate rotations about
the $u$, $v$, and $w$ axes, respectively.  

In addition, local $u'$ and $v'$
coordinates are defined such that they are parallel to $u$ and $v$, but the
direction is always chosen to be in positive $\phi$, $z$, or $r$ directions,
irrespective of the orientation of the local coordinate system. For the TID
and TEC wedge-shaped sensors, where the topology of the strips is radial, the
$u'$- and $v'$-axes change direction across the sensor such that $v'$ is
always directed along the strips and therefore $u'$ corresponds to the global
$r\phi$-coordinate.

\section{Optical Survey Measurements}
\label{sec:survey}

Optical surveys taken during module construction and integration
provide initial alignment parameters for many of the modules.
Additionally, the survey information was used as a constraint in the
alignment procedure, as will be described in Section \ref{sec:quality}.

Coordinate Measuring Machine (CMM) data and photogrammetry have been
used for the optical survey of the tracker components.  
While the former were used for measurements of the active elements,
the latter were used for the alignment of larger structures. 
For the inner strip detectors (TIB and TID), both
module-level and high-level structure information were used in the analysis.
For the outer strip detectors (TOB and TEC),
module-level survey data were used only as a conformance criterion during construction, 
while survey information from high-level structures was used in the analysis.  More
information on the strip detector surveys can be found in Ref.~\cite{tifalignment}.

Detailed optical surveys of the pixel endcap detectors were performed
as part of the construction process. First, module positions were measured within a panel,
which contains three or four modules. Then the positions of modules were
also measured on a half-disk, where 12 panels are placed on each
side. Eight fiducial points were measured on the active area of each module, 
allowing a position measurement precision of about one micron. 
Some of the module fiducial points were partially obscured by other modules 
when mounted on a half-disk and thus could not be measured. 
Good reproducibility of module measurements within a panel 
was found in the two configurations, allowing an estimate
of survey precision. Measurements of the two sides of the half-disk
were connected through a touch-probe, which related positions 
of the touch-probe targets previously measured with CMMs.
This also related measurements from the front and the back panels of blades.
Finally, half-disks were placed in the half-cylinders,
which were then inserted in the pixel detector volume. 
Half-disk and half-cylinder positions were measured with photogrammetry 
and were related to the positions of the active elements through photo targets, 
which had been previously measured with CMMs.

In the barrel pixel detector, only two-dimensional measurements of the module
positions within a ladder were performed. High-resolution digital images of
four nearby fiducial points in neighboring modules were taken. A
precision of 2 $\mu$m  was achieved for longitudinal positions of the modules
within a ladder, allowing, in particular, a tight constraint on the $z$-scale. 
The relative position of nearby ladders in the half-shells were also surveyed
with similar two-dimensional measurements, but were not used in this analysis.


\section{Track-based Alignment}
\label{sec:stat}

The goal of the track-based alignment procedures is to determine the module positions
from a large sample of reconstructed charged particle
trajectories. Each trajectory is built from charge depositions (``hits'') on
individual detectors, assuming a piece-wise helical track model, incorporating
effects from multiple scattering and energy loss. The ``Combinatorial Track
Finder'' (CTF) track algorithm~\cite{crafttracking} was
used to reconstruct the cosmic muon trajectories. 
The result is a set of track parameters.
Five track parameters describe the helical trajectory of a track at the 
point closest to the nominal interaction point:
distance of closest approach in the transverse $d_{xy}$ and
longitudinal $d_z$ directions,  
track azimuthal angle $\phi$,
track polar angle $\theta$,
and transverse momentum $p_T$. 

\begin{figure}[bh]
\begin{center}
\includegraphics*[angle=0,width=1.00\linewidth]{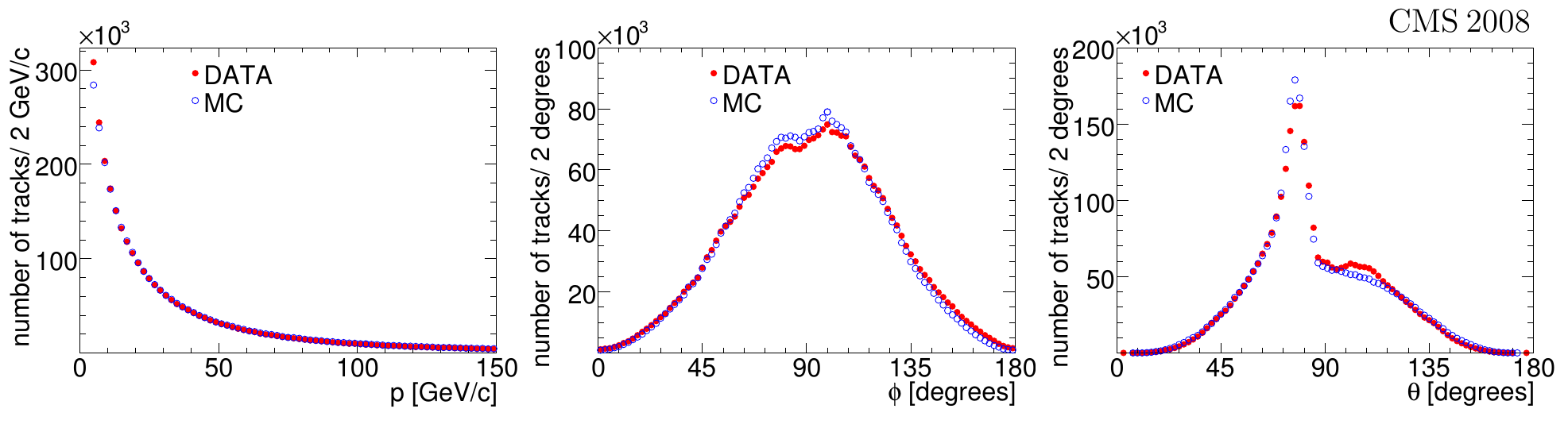}
\caption{
Momentum (left),  azimuthal angle (middle), and polar angle (right) spectra
of cosmic muons reconstructed in the CMS tracker volume
based on the selection criteria described in the text.
The solid (red) circles represent the cosmic ray data whereas the open (blue) circles come from a MC simulation.
The vertical cosmic ray tracks correspond approximately to $\phi=\theta=90\degree$.
}
\label{fig:momspectrum}
\end{center}
\end{figure}

The momentum and angular spectra of cosmic muons reconstructed in the CMS tracker volume
are shown in Fig.~\ref{fig:momspectrum}, where good agreement with MC simulation~\cite{CMSCGEN} 
is observed. These track parameters are defined at the point of closest approach to the CMS nominal 
interaction point and the absolute value of the $\phi$ angle is shown, considering all particles traversing 
CMS from top to bottom. The aligned tracker geometry was used in track reconstruction.
The asymmetry about $\phi=90^\circ$ is due to the excess of positive over negative cosmic ray 
particles entering CMS and the bending of charged tracks in the magnetic field of the solenoid. 
Positively charged cosmic rays tend to peak just above $90^\circ$ whereas negatively charged 
cosmic rays tend to peak just below $90^\circ$. The asymmetry about $\theta=90^\circ$ is mainly 
due to the asymmetric location of the cavern shaft.
Tracks were selected according to the following criteria. 
Each track was required to consist of at least eight hits with a
signal-to-noise ratio larger than 12 in the silicon strip modules
or a probability of the pixel hit matching the template shape~\cite{templates} of at least
0.001 (0.01) in the local $u$ ($v$) direction.
Hits were also rejected if the track angle relative to the local $uv$  
plane was less than $20\degree$. 
In addition, two hits were required to be on either pixel or double-sided strip modules,
allowing a precise measurement of the polar angle $\theta$.
The particle momentum had to be greater than 4\,GeV/$c$, and 
the $\chi^2$ per degree of freedom of the track fit, $\chi^2/\mathrm{ndf}$, had to be less than 6.0.
Outlying hits with large normalized residuals were removed.
In total, about 3.2 million tracks were selected for alignment, out of which
about 110\,000 had at least one pixel hit.

The track reconstruction used the Alignment Position Errors (APE),
the estimated uncertainty on the module position in the three global coordinates,
which were added in quadrature to the hit errors during the pattern
recognition and track fitting procedure. This allowed efficient track
reconstruction in the presence of misalignment and a correct pull
distribution of track parameters.
The determination of APEs is described in Section~\ref{sec:ape}. 
The APE values used for the initial track reconstruction before alignment analysis
were set to large values as they had to account for possible large displacements 
of the entire sub-detectors while still guaranteeing an efficient track-hit association.

\subsection{Alignment algorithms}
\label{sec:quality}

The alignment is an optimization problem that can be formulated in the
context of linear least squares. Module position corrections (``alignment
parameters'') ${\bf p}$ are determined by minimizing an objective function
\begin{equation}
\label{eq:chisq}
\chi^2({\bf p},{\bf q})
= \sum_j^{\rm tracks}\sum_i^{\rm hits}{\bf r}_{ij}^T({\bf p},{\bf q}_j){\bf V}^{-1}_{ij}{\bf r}_{ij}({\bf p},{\bf q}_j)\,,
\end{equation}
which can be expressed  as the sum over all hits $i$ on all tracks $j$ and track parameters ${\bf q}_j$, 
assuming negligible correlations between hits. Track residuals ${\bf r}_{ij} = 
{\bf m}_{ij} - {\bf f}_{ij}({\bf p},{\bf q}_j)$ are defined as the difference
between the measured hit position ${\bf m}_{ij}$ and the trajectory impact
point ${\bf f}_{ij}$, both of which are given in the module local coordinate
system. The residuals are either one- or two-dimensional vectors,
depending on the type of module, and ${\bf V}_{ij}$ is either the
squared error or the covariance matrix in case of one- or
two-dimensional residuals, respectively. 

Two statistical methods were employed to solve the alignment problem. 
Both of them were previously applied to the CMS silicon strip tracker 
alignment  during stand-alone commissioning~\cite{tifalignment}.
The global alignment algorithm (``Millepede~II'')~\cite{MillepedeII} minimizes the
$\chi^2$ function in Eq.~(\ref{eq:chisq}) by taking into account track and
alignment parameters simultaneously. 
This algorithm has also been previously used in simulation studies of the CMS tracker~\cite{mpmc}.
Since only the $n$ alignment parameters ${\bf p}$ are of interest, 
the problem is reduced to the solution of a matrix
equation of size $n$. The value of $n$ is of the order of $10^5$ for
six degrees of freedom of 16\,588 modules.
The covariance matrix ${\bf V}_{ij}$ is approximated by a
diagonal matrix with uncertainties $\sigma_{ij}$ for uncorrelated hit
measurements ${\bf m}_{ij}$ of the track $j$.
Given reasonable starting values, ${\bf p}_0$ and ${\bf q}_{j0}$, 
the track model prediction ${\bf f}_{ij}({\bf p},{\bf q}_j)$ can be
linearized. Since angular corrections are small, the linearized problem is a
good approximation for alignment.

The local iterative algorithm (``Hits and Impact Points'')~\cite{hipalignment}
approximates Eq.~(\ref{eq:chisq}) by assuming no track parameter ${\bf q}$ dependence
and therefore ignores correlations between alignment parameters for different
modules in one iteration. 
The trajectory impact point ${\bf f}_{\alpha j}$ is recalculated 
for each hit, removing the hit under consideration from the track fit. 
The track parameters and correlations between different modules are taken into account
through iterations of the minimization procedure and refitting the tracks with
new alignment constants after each iteration. 
The number of iterations is determined empirically.
This approach allows a simplification of the formalism and
leads to a solution of the six-dimensional matrix equation for the six
alignment parameters of each module in each iteration.

The local iterative algorithm permits the inclusion of survey measurements 
in the formalism of Eq.~(\ref{eq:chisq}), as described in Ref.~\cite{babaralignment}. 
This leads to an additional term in the objective function 
to be minimized independently for each module in a given iteration:
\begin{equation}
\label{eq:chisqsurvey}
\chi_m^2({\bf p})
= \sum_{i}^{\rm hits}{\bf r}_{i}^T({\bf p}){\bf V}^{-1}_{i}{\bf r}_{i}({\bf p})
+\sum_{j}^{\rm survey}{\bf r}_{*j}^T({\bf p}){\bf V}^{-1}_{*j}{\bf r}_{*j}({\bf p})\,,
\end{equation}
where the original formulation has been modified to include survey
information from the $k$ hierarchial levels for each sub-detector.
The track residuals ${\bf r}_{i}({\bf p})$ do not have explicit dependence on track 
parameters and enter the sum over hits in a given module $m$.
The six-dimensional survey residuals ${\bf r}_{*j}$ are defined as the
difference between the reference and the current sensor position. The survey
measurement covariance matrix ${\bf V}_{*j}$ reflects both the survey
precision and additional uncertainties due to changes in the detector. These
errors can be configured differently for different hierarchy levels and
for the degrees of freedom that should be stable, such as the longitudinal
direction in a barrel ladder, and those which may change more with time.

The local iterative method uses the
full implementation of the Kalman filter track reconstruction algorithm adopted
in CMS~\cite{crafttracking}. It requires a large number of iterations and large computing
resources to refit the tracks in each iteration. The global
method, instead, allows the determination of alignment parameters, properly
accounting for the correlations among them, in a single step. 
However, the global method does not take into account the effects of
material in the tracker and assumes a simple helical trajectory for
charged particles. 
The method also requires a large amount of computer
memory and the application of advanced techniques~\cite{MillepedeII} 
for solving Eq.~(\ref{eq:chisq}).
Each of the two alignment algorithms was used to obtain module positions
independently and a comparison of results between the two complementary 
approaches was part of the validation of the tracker alignment procedure. 
After verifying that the two methods yielded consistent results, the final results 
were obtained by applying the two algorithms in sequence in order to take 
advantage of their complementary strengths.

\subsection{Alignment strategy}
\label{sec:doublesided}

Single-sided silicon strip modules provide only a one-dimensional measurement 
in the module plane, along the local $u$-coordinate, which effectively translates 
into a two-dimensional measurement in space. The $v$-coordinate is only known 
to be within the module boundaries, with precision not sufficient for track 
reconstruction requirements. However, the information from the $r\phi$ and 
stereo modules in a double-sided module is combined into a two-dimensional 
measurement in the combined module plane in both $u$ and $v$ for the pattern 
recognition phase. 

Due to the 100 mrad stereo angle between the $r\phi$ and stereo modules, though,
a small displacement in $u$ is equivalent to a ten times larger displacement in $v$. 
Given comparable mounting precision of modules in $u$ and $v$, a smaller 
$u$ displacement is much more likely than a larger $v$ displacement. In fact, it was 
found from the data after several attempts at alignment of double-sided modules 
in $v$, that parameters much larger than the known assembly accuracy were obtained.  
Whenever possible, the two single-sided components of a double-sided module were 
aligned separately, but only in the most precise coordinates. The relative  alignment 
of the two sides was found to be consistent with assembly tolerance and at the same 
time improved track residuals significantly. This fact and the hierarchical structure
of the tracker led to multi-step strategies for both alignment methods.

\subsection{Alignment with the global method}
\label{sec:mp}

The first step of the global method aligns
the highest-level structures (half-barrels, endcaps) with all six degrees of freedom 
together with all module units, including $r\phi$ and stereo strip modules
in a double-sided module,  with the most sensitive degrees of freedom each
($u$, $w$, $\gamma$, and for pixel modules also $v$). 
The detector design geometry was chosen as a starting point.
A limitation of 46\,340 parameters in the program used led to selection
criteria for modules to be aligned of more than 25 hits per pixel module
and more than 425 hits per strip module.  This resulted in the alignment of
about 90\% of all modules.

In the second step all modules (all double-sided or single-sided strip modules 
with more than 150 hits and all pixel modules with more than 25 hits) 
were aligned in the TIB in $u, w, \alpha, \beta, \gamma$; in the
pixel system in $u, v, w, \gamma$; and in $u, w, \gamma$ elsewhere.
Compared to the previous step, this recovered the remaining 10\% of the modules
and allowed more degrees of freedom for the TIB, which had larger assembly tolerance,
but did not allow independent alignment of the $r\phi$ and stereo modules within double-sided combinations.
No alignment of $\alpha$ and $\beta$ was performed in the TOB due to its higher mounting accuracy, 
and in the TID and TEC, since they were less often traversed by cosmic ray tracks.

The third step was designed to recover lost correlations between the first two steps
and had the same configuration as the first step, but the minimum number of
hits in the strip modules was increased to 450 with respect to 425 used in the first step.

\subsection{Alignment with the local method}
\label{sec:hip}

Module positions as determined from optical surveys were used
as the starting point in the local method. The difference of the starting geometry
to that of the global method was motivated by the use of survey information
in the minimization procedure, as described in Section~\ref{sec:quality}. 
However, for both methods the initial geometry did not affect the results 
significantly. The alignment with the local algorithm started with the large 
structures and then proceeded in order of increasing granularity
down to the module units in order to speed up convergence. 
The small fraction (4\%) of tracks passing through the pixel detector
suggested a modified approach in which the silicon strip modules were
first aligned without pixel information.
The hits from the structures not yet aligned were excluded from the track fit.
The  APE values were set high at the beginning of the alignment process 
(several hundreds of microns depending on the
sub-detector), and were progressively reduced to a few tens of microns
in subsequent alignment steps.
The values of the APEs were kept fixed within each alignment step.

The alignment with the local method was carried out in six steps.  
In the first step (15 iterations), the half-barrels of BPIX, TIB, and TOB were  
aligned as rigid bodies with four degrees of freedom ($u$,$v$,$w$,$\gamma$). 
In order to have a better sensitivity along the global $z$-direction,
only tracks passing through the pixel barrel system were used. 
In the second step (15 iterations), the endcaps of the FPIX,
TID, and TEC were aligned in the same four degrees of freedom.
The third alignment step (30 iterations) moved the strip modules (treating 
double-sided modules as rigid bodies)  in all six degrees of freedom. 
Data from the optical survey measurements (or design geometry if no survey
information was available) were used according to
Eq.~(\ref{eq:chisqsurvey}). The information on the module positions
within a higher-hierarchy structure, for example within a string of TIB, was
used in the $\chi^2$ minimization.
This proved to be useful for both limiting the movement of modules in
poorly constrained degrees of freedom and aligning modules that had
fewer than the required number of hits.
In the fourth step (30 iterations), the strip module units
(treating $r\phi$ and stereo strip modules in a double-sided module independently)
with at least 50 hits were aligned in three degrees of freedom 
($u$, $w$, $\gamma$).
In step five (15 iterations), 
the ladders of the pixel detector were aligned in six degrees of freedom. 
Finally in the sixth step (15 iterations), the pixel modules with at least 30 hits
were aligned in all six degrees of freedom.

After every step the alignment parameters were checked for convergence. Fewer than 20 modules failed to converge since they
were moved to implausible positions by the algorithm. 
The alignment parameters for these modules were left at the survey values.
It was found that in some cases the basic procedure needed to be
applied a second time in order to improve the convergence for the alignment of the large-scale structures. 
However, as will be discussed in the next section, the final
alignment results were derived from a sequential approach involving
both the global and the local method, and for this reason the local
method was limited to a single set of steps.

\subsection{Alignment with the combined method}
\label{sec:superhipmp}

The strength of the global method is solving effectively the global
correlations, while for the local method it is that
the same track fit is used as in the standard CMS reconstruction,
and that survey information can easily be incorporated 
thus allowing for alignment with more degrees of freedom.
In order to take advantage of both methods, the final alignment parameters were
produced starting from the output of the global method analysis
described in Section~\ref{sec:mp}, then further aligning the tracker with the
local method presented in Section~\ref{sec:hip}. 

The alignment strategy adopted by the local algorithm was modified to 
exploit the already good starting position of the modules provided 
by the global method. 
The APEs were kept fixed at 10\,$\mu$m in the barrels and 30\,$\mu$m in
the endcaps, similar to those described in Section~\ref{sec:ape}.
In the first step of the local method (30 iterations), all strip modules
(treating double-sided modules as rigid bodies)
were aligned in six degrees of freedom using track and survey information.
In the second step (20 iterations), the strip module units
(treating $r\phi$ and stereo strip modules in a double-sided module independently)
were aligned in three degrees of freedom ($u$,$w$,$\gamma$).
Pixel modules were not aligned in the first two steps, although pixel hits were included in the track fit. 
Finally, in the last step (20 iterations), the pixel modules
were aligned in six degrees of freedom.
Modules for which the fit did not converge were left at the position obtained by the global algorithm.


\section{Results and Validation}
\label{sec:valid}

The alignment analysis was performed after the proper calibration of 
both the strip and pixel detectors~\cite{crafttracking, craftpixel}.
In particular, an appropriate Lorentz angle calibration was applied.
A Lorentz angle miscalibration would result in a systematic shift 
of the effective charge collection position, 
in the same $u$-direction for all the modules with the same
relative orientation between the solenoid and the
drift field.
Since this would appear as an effective shift of the module, 
it would not be evident if corrected for by the alignment.
Therefore, for validation purpose only, alignment results were checked with
the data collected both with 0\,T and with 3.8\,T solenoid fields. 
The precision achieved in the alignment analysis is such that there 
should be clear evidence for a Lorentz angle miscalibration with an
effective shift of only a few microns. In fact, good agreement with the
Lorentz angle calibration~\cite{crafttracking} was found. 

Several approaches were employed to validate the alignment results.
The low-level quantities that were used in the $\chi^2$ minimization, 
such as residuals and the $\chi^2/\mathrm{ndf}$ of the tracks, were monitored.
Given the limited size of the alignment track sample, it was used both
for alignment and validation. In addition, a high-level   
validation analysis was performed to monitor the track parameter resolution. 
Moreover, techniques for comparing positions of modules within
differently derived geometries, described in
Sections~\ref{sec:systematics} and~\ref{sec:geometry},
allowed a better understanding of the alignment performance.

\subsection{Calibration of alignment position errors}
\label{sec:ape}

\begin{figure}[tbh]
\begin{center}
\includegraphics*[width=\linewidth]{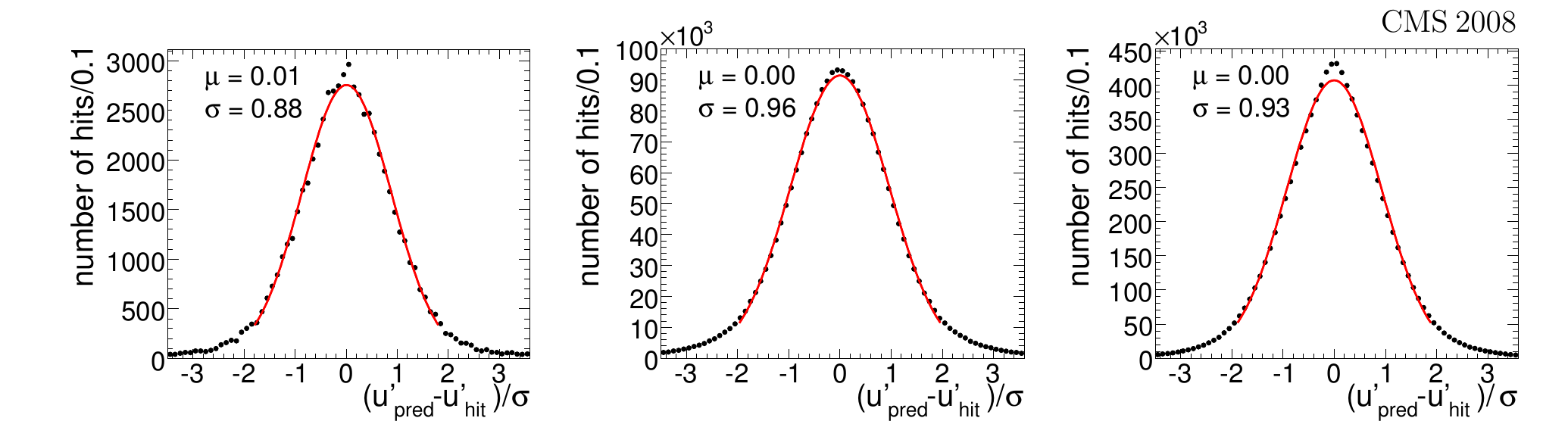}
\caption{
Distributions of normalized track residuals after the APE calibration
procedure: BPIX (left), TIB (center), and TOB (right). Solid lines
represent the results of Gaussian fits and the fit mean and sigma
values are given within the plots. The 
alignment obtained from the combined method was used.  }
\label{fig:APE}
\end{center}
\end{figure}

In the track refit the APEs for each hit were assumed to be the same in the 
three spatial directions.
For an alignment performed using cosmic ray tracks, the precision of the
alignment varies within a given sub-detector because of the different
illumination of modules due to their orientation relative to cosmic rays.
The radius of the sphere representing the APE for each module, $r_{\rm sphere}$, was therefore
taken to be $r_{\rm sphere}={r_0}/{\sqrt{N_{\rm entries}}}$,
where $N_{\rm entries}$ represents the number of hits in a module.
The value of $r_0$ was chosen in order to have the Gaussian standard deviation 
of the distribution of the residuals normalized to their error approximately equal to one 
in the symmetric interval covering 95\% of the distribution.
In the TIB and TOB, a common $r_0$ value was defined for each layer and
separately for the $r\phi$ and stereo components of the double-sided layers. 
Elsewhere the value of $r_0$ was defined at the sub-detector level.
The values of $r_{\rm sphere}$ were restricted to reasonable values, especially in the
case of small or zero $N_{\rm entries}$, where a precision 
compatible with survey and assembly data was used.
APEs were calibrated using the set of alignment parameters obtained from
the combined method. Examples of the distribution of the normalized
residuals after the calibration of the APEs are shown in Fig.~\ref{fig:APE}.

\subsection{Track fit quality and hit residuals}
\label{sec:residuals}

\begin{figure}[tbh]
\begin{center}
\includegraphics*[width=0.5\linewidth]{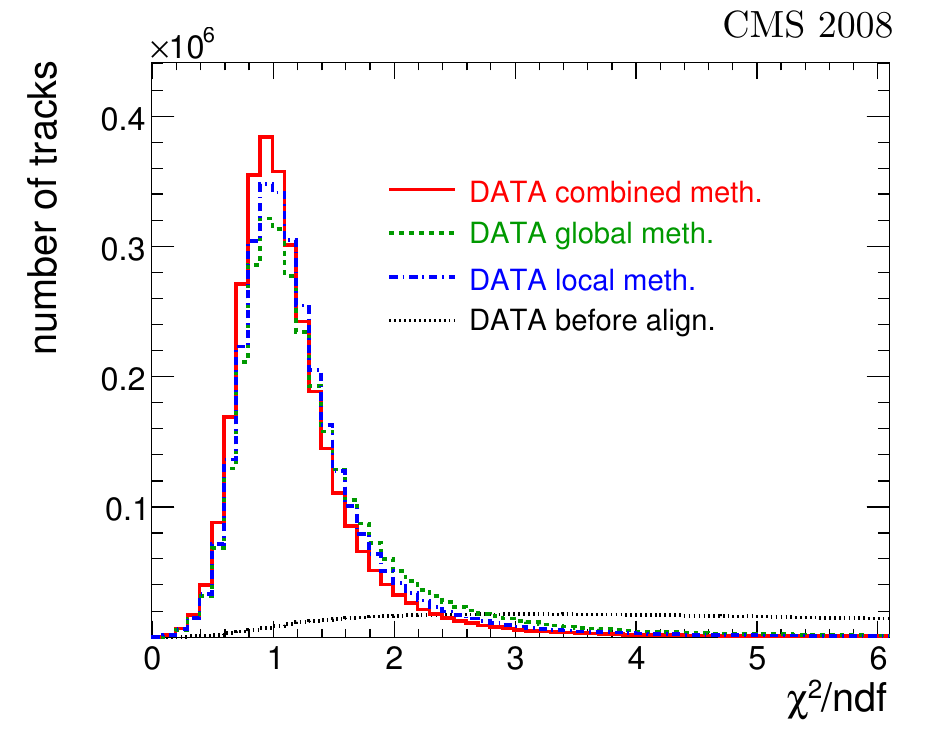}
\caption{
Distributions of the $\chi^2/\mathrm{ndf}$ of the tracks
before alignment (dotted line) and after alignment with the
local (dashed-dotted line), 
global (dashed line), 
and combined (solid line) methods.
}
\label{fig:chi}
\end{center}
\end{figure}

\begin{figure}[tbh]
\begin{center}
\includegraphics*[width=0.9\linewidth]{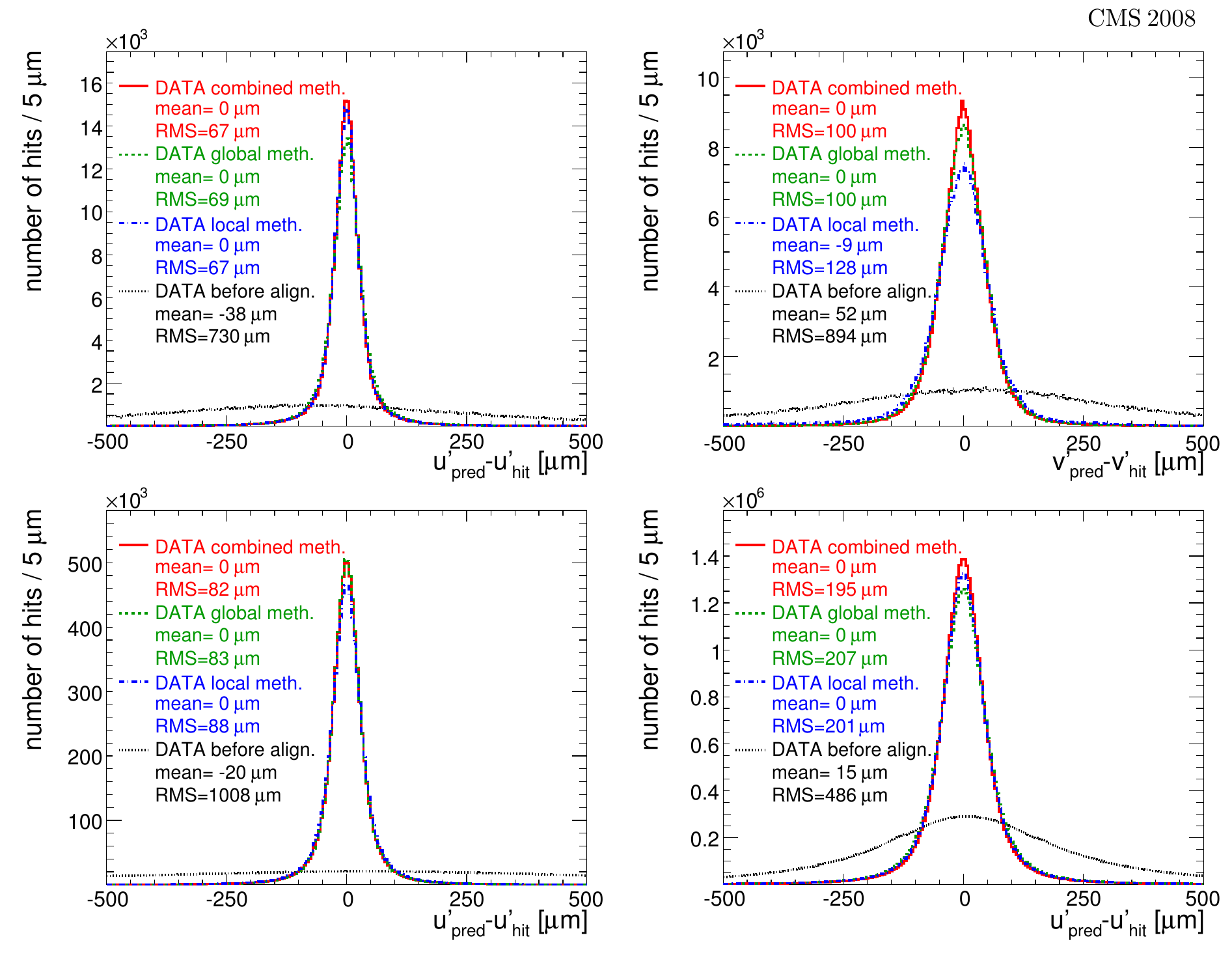}
\caption{
Track residuals, shown for BPIX (top left $u'$, top right $v'$),
TIB (bottom left), and TOB (bottom right).
The four lines correspond to positions before alignment (dotted lines) and after alignment 
with the global (dashed lines), local (dot-dashed lines), and combined methods (solid lines). 
}
\label{fig:resid}
\end{center}
\end{figure}

\begin{figure}[ptbh]
\begin{center}
\includegraphics*[width=0.9\linewidth]{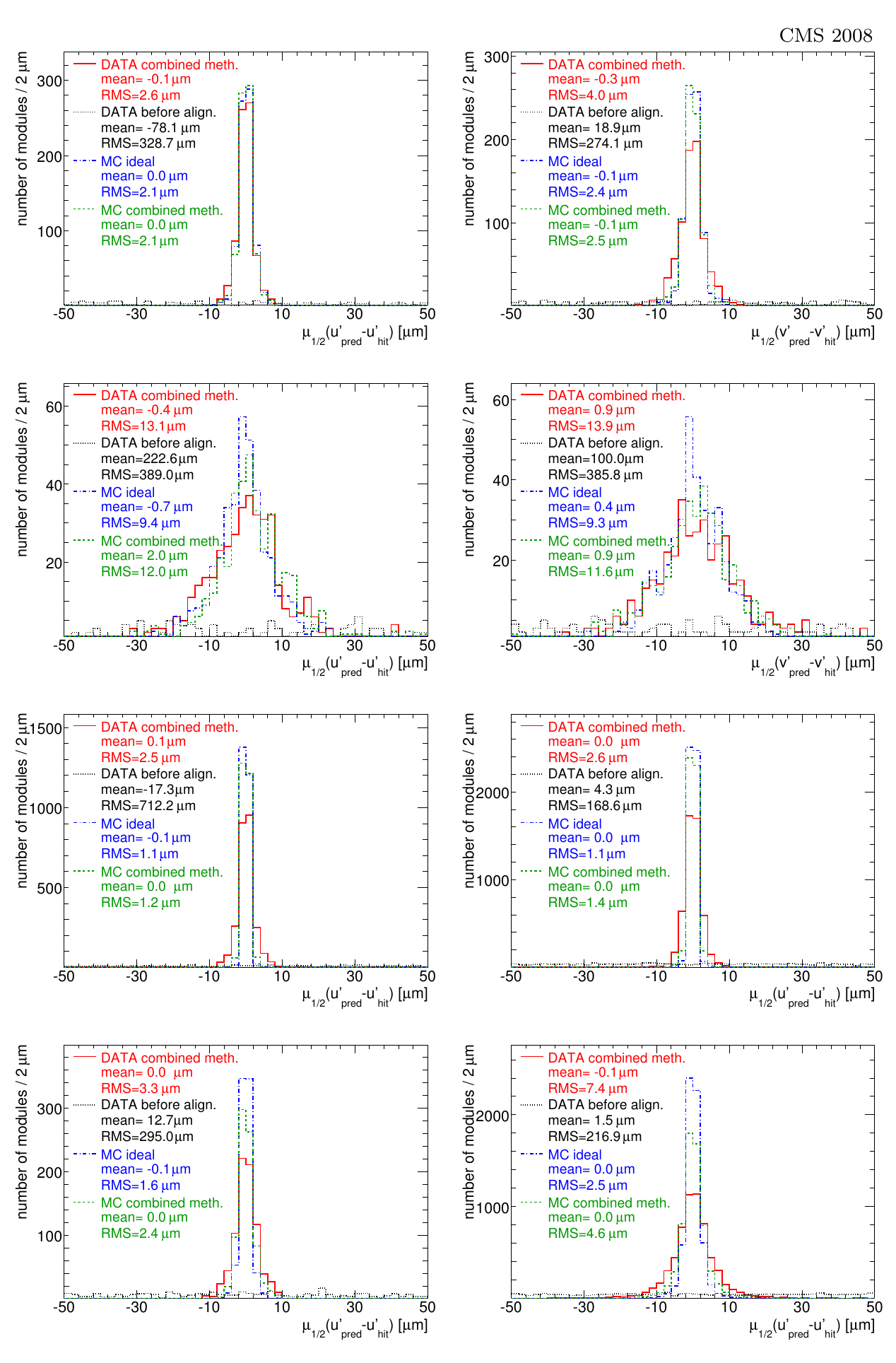}
\caption{
Distribution of $\mu_{1/2}$, the median of the residuals, for modules 
with more than 30 hits, shown for BPIX (top left $u'$, top right $v'$),
FPIX (second row left $u'$, second row right $v'$),
TIB (third row left), TOB (third row right),
TID (bottom left), and TEC (bottom right).
Shown are distributions before alignment (black dotted), after alignment with the combined method (red solid), combined method MC (green dashed), and ideal MC (blue dash-dotted).
}
\label{fig:dmr}
\end{center}
\end{figure}

All tracks used in the validation procedure were refitted with APEs corresponding
to the alignment parameters obtained with the combined method.
The loose track selection described in Section~\ref{sec:stat} was applied.
To avoid a bias in the determination of the hit residuals, 
the track prediction was calculated without information from the hit under consideration. 
The track $\chi^2/\mathrm{ndf}$ distribution is shown in Fig.~\ref{fig:chi}.
The hit residuals in the $u'$ and $v'$ directions are shown in Fig.~\ref{fig:resid}.
The hit residual width is dominated by two effects other than alignment: 
track extrapolation uncertainties due to multiple scattering 
and hit position reconstruction uncertainties. 
Both of these effects are random, while misalignment leads to systematic
shifts of the residuals. For this reason, the distribution of the
median of the residuals (DMR) is taken as the most
appropriate measure of alignment. Median distributions are shown in
Fig.~\ref{fig:dmr} and the corresponding RMS values of these
distributions are given in Table~\ref{tab:dmr}. 

To check the statistical precision of track-based alignment a MC simulation
was performed in which module positions from the combined method obtained
with data were used as the starting geometry in the MC alignment procedure.
This approach in MC effectively models the situation in data prior to and
during the alignment. The resulting DMRs are also shown in Fig.~\ref{fig:dmr} 
and the RMS values listed in Table~\ref{tab:dmr}. 
For comparison, the distributions obtained 
from the ideal MC simulation are presented 
in Fig.~\ref{fig:dmr}  as well.

The middle columns of Table~\ref{tab:dmr} reflect different stages of alignment:
the initial situation with no alignment, alignment with tracks using the
global and local methods, and finally with the combined method. 
Table~\ref{tab:dmr} also includes results from MC simulations based on
combined and ideal geometries. For the 2D pixel detectors both $u'$ and $v'$ coordinate measurements are quoted. 
Overall, there is significant improvement in the
track reconstruction going from the geometry without any alignment,
to the alignment using tracks with the local and the global method, 
and finally to the combined result. With respect to cosmic ray trajectories,
the module positions were determined to an average precision of  3--4\,$\mum$ RMS 
in the barrel and 3--14\,$\mum$ RMS in the endcaps in the most sensitive coordinate.
These values are in agreement with the expected
statistical precision as determined using simulated events. 
They are also comparable to values obtained from a MC  
simulation based on the ideal detector geometry.
\begin{table*}[tbh]
\caption{
RMS of the distribution of the median of the residuals (DMR)
in the $u'$ and $v'$ local coordinates for modules with more than 30 hits.
The number of these modules compared to the total number of
modules is stated in the last column. Four geometries are considered: 
those obtained with the three methods discussed in the text and the 
geometry before alignment. Results from simulations based on the combined 
alignment and ideal geometries are shown for comparison.
}
\label{tab:dmr}
\begin{center}
\begin{tabular}{|c|c|c|c|c|c|c|c|}
\hline
      &   before &  global & local & combined &  combined  & ideal & modules \cr
      &    [$\mu$m]      &  [$\mu$m]      & [$\mu$m]      & [$\mu$m]         &  MC    [$\mu$m]   &  MC    [$\mu$m] & $>$30 hits \cr
\hline
BPIX  ($u'$) &   328.7  &  7.5 &  3.0 &   2.6   &   2.1 &  2.1 &  \cr
BPIX  ($v'$) &   274.1  &  6.9 & 13.4 &   4.0   &   2.5 &  2.4 &  \raisebox{1.5ex}[-1.5ex]{757/768} \cr
FPIX  ($u'$) &   389.0  & 23.5 & 26.5 &  13.1   &  12.0 & 9.4 &  \cr
FPIX  ($v'$) &   385.8  & 20.0 & 23.9 &  13.9   &  11.6 & 9.3 &  \raisebox{1.5ex}[-1.5ex]{393/672} \cr
TIB  ($u'$) &   712.2  &  4.9 &  7.1 &   2.5   &   1.2 & 1.1 &   2623/2724  \cr
TOB  ($u'$) &   168.6  &  5.7 &  3.5 &   2.6   &   1.4 & 1.1 &  5129/5208 \cr
TID  ($u'$) &   295.0  &  7.0 &  6.9 &   3.3   &   2.4 &  1.6 &  807/816   \cr
TEC  ($u'$) &   216.9  & 25.0 & 10.4 &   7.4   &   4.6 &   2.5 & 6318/6400 \cr
\hline
\end{tabular}
\end{center}
\end{table*}

\clearpage

\subsection{Residuals in overlapping module regions}
\label{sec:overlaps}

A further method to monitor and validate the results of the alignment is
to use the hits from tracks passing through regions where modules
overlap within a layer of the tracker.
This method, described in detail in Ref.~\cite{tifPaper},
is also used to measure the hit resolution of the sensors~\cite{crafttracking}.
In this method, the difference in residual values 
for the two measurements in the overlapping modules is compared, once the
hits in the layer under consideration are removed from the track fit.
The proximity of the hits reduces the amount of material between the two 
hits and minimizes the uncertainties on the predicted positions associated with track propagation.
Deviations between the reconstructed hits and the predicted positions
allow an assessment of the relative alignment between two adjacent
modules.
A non-zero mean of a Gaussian fit to the distribution of residual
differences indicates a relative shift between the modules. 
A relative rotation between the modules can be seen in the dependence
of the differences between the residuals and the track position predictions
on the modules under study, where the slope indicates the
magnitude of the residual rotation.
This would also lead to an increase in the width of the Gaussian fitted 
to the distribution of the differences between the
residuals, and can be minimized by requiring that the slopes be small, 
as is done in the hit resolution measurement~\cite{crafttracking}.
Any quantitative assessment is nevertheless difficult since it is not easy
to disentangle the contributions of the rotations around the different
axes. 

Only events with a single reconstructed track were used in this analysis.
Module pairs with more than 35 (100) hits in the overlap region were 
analyzed in the pixel (strip) detectors. Only barrel layers satisfied these 
requirements. The lower hit requirement in the pixel modules reflects their 
smaller area and leads to correspondingly less precise mean values.

\begin{figure}[htb]
\begin{center}
\includegraphics*[width=0.9\linewidth]{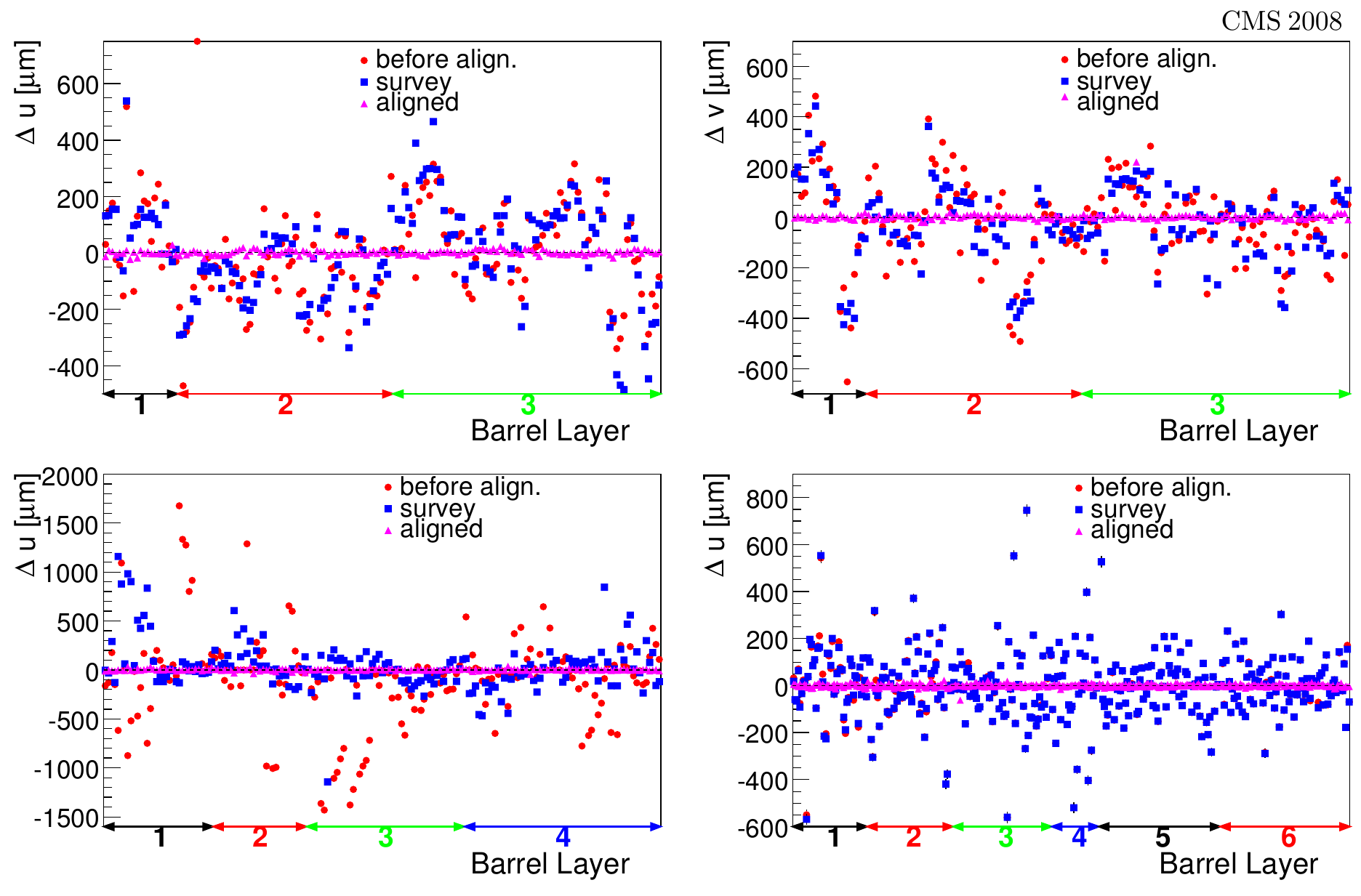}
\caption{
Relative shift between module pairs in the local $u$ (top left) and local
$v$ (top right) direction in the BPIX, 
in the local $u$-coordinate in the TIB (bottom left),
and TOB (bottom right). 
Only modules in the slice
$80^\circ < \phi<100^\circ$ are shown in the TIB and TOB.
}
\label{fig:overlap_pxb}
\end{center}
\end{figure}

\begin{table*}[htb]
\caption{
RMS of the mean of the distributions of the relative shift between overlapping module pairs,
scaled by $1/\sqrt{2}$ to account for the two independent measurements. 
Results are shown for the three data and two MC geometries described in the text.
}
\label{tb:overlap_rms}
\begin{center}
\begin{tabular}{|c|c|c|c|c|c|}
\hline
     &   before &  survey & combined & combined & ideal \\
     &    [$\mu$m]     &  [$\mu$m] &  [$\mu$m] & MC  [$\mu$m]  & MC  [$\mu$m] \\
\hline
BPIX  ($u$) 	& 114	& 121	& 5.7  & 4.3 & 4.4 \\
BPIX  ($v$) 	& 122	& 110	& 12.7 & 4.7 & 4.2 \\
TIB  ($u$)    	& 264	& 187	& 7.0 &  1.6 & 1.1  \\
TOB  ($u$) 	& 118	& 118	& 5.1 &  2.1 & 1.6  \\
\hline
\end{tabular}
\end{center}
\end{table*}
The relative shifts are shown in Fig.~\ref{fig:overlap_pxb}. 
In the strip detectors, the relative shift can be measured in the local $u$-coordinate.
In the pixel detector, this shift can be measured both in the local $u$ and
local $v$ coordinates.
The RMS of the distribution of the mean values are given in Table~\ref{tb:overlap_rms}, where
the values are scaled by $1/\sqrt{2}$ to account for two independent measurements. 
The corresponding mean values of the distributions after alignment are 1.5~$\mu$m or less.
The expectation from MC, for both aligned as well as ideal geometry, is in good agreement 
with the distribution of the median of residuals quoted in Table~\ref{tab:dmr}, where
results in the pixel detector are more affected by the small size of the overlap sample.
The differences in data may be a sign of yet unquantified systematic effects, 
like the aplanar distortions of modules.
While the distributions of hit residuals taken over the full
active region of the modules tend to be well centered, this is not  
always the case for distributions in the overlap regions, which are  
located at the edges of the active areas in the modules.
However, this effect is smaller then 10\,$\mu$m in the most precise 
$u$ direction and is 13\,$\mu$m in the $v$ direction in the pixel barrel modules.
There is also visible improvement of the results with survey geometry compared to the
geometry before alignment in the TIB, while in the TOB no module-level survey was done and
in the BPIX no survey data between overlapping ladders were used either.

\subsection{Track parameter resolution}
\label{sec:split}

\begin{figure}[htb]
\begin{center}
\includegraphics*[width=0.9\linewidth]{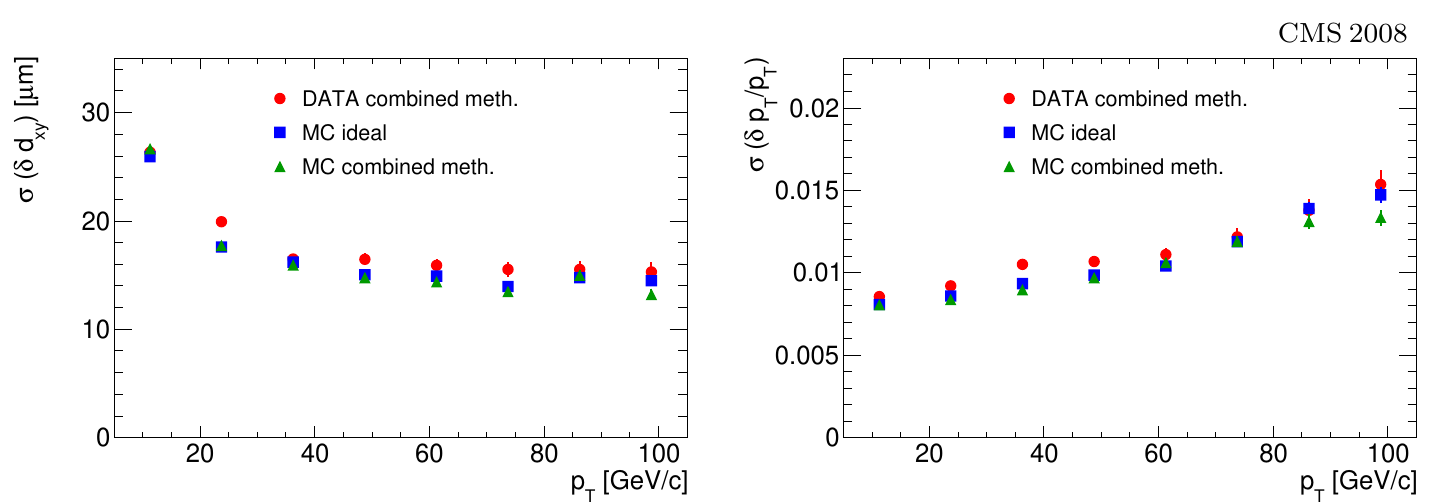}
\caption{
Dependence on $p_T$ of the differences between the track parameters measured  
at the point of closest approach to the nominal beamline, in the two
halves of a cosmic ray track and scaled by 1/$\sqrt{2}$.
The RMS of the distribution truncated at 95$\%$ is quoted at each momentum interval, 
shown for the distance of closest approach in the transverse direction $d_{xy}$ (left) 
and for $1/p_T$ (right). 
Results are shown for the geometries derived from ideal MC (blue squares),
and from the alignment result with cosmic ray data (red circles) 
and combined method MC (green triangles).
}
\label{fig:split-vspt}
\end{center}
\end{figure}

The track parameter resolutions can be validated with 
independent reconstruction of the upper and the lower portions of
cosmic ray tracks and comparison of track parameters at the point of 
closest approach to the nominal beamline. 
Both the upper and lower track segments were required to have at least three pixel hits.
This mimics the topology of collision tracks.
\begin{figure}[htb]
\begin{center}
\includegraphics*[width=0.9\linewidth]{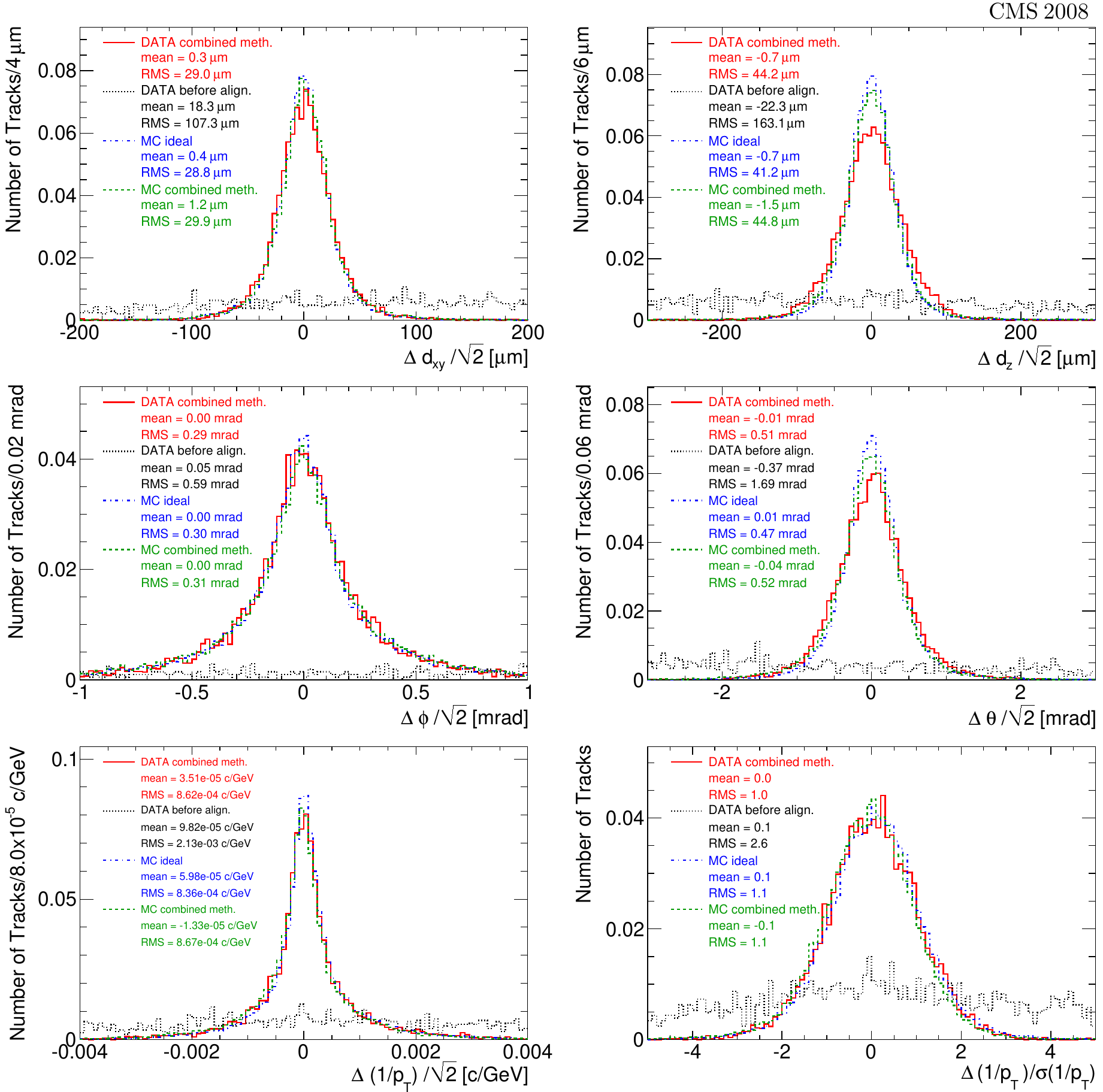}
\caption{
Differences between upper and lower track segment parameters measured 
at the point of closest approach to the beamline and scaled by $1/\sqrt{2}$.
Distributions are shown for  the distance of closest approach in 
the transverse direction $d_{xy}$ (top left),
the same in the longitudinal direction $d_z$ (top right),
the track azimuthal angle $\phi$ (middle left), 
the track polar angle $\theta$ (middle right), 
and $1/p_T$ (bottom left). 
The plot on the bottom right shows the $1/p_T$ difference normalized to its error, that is
$(1/p_{T,1}-1/p_{T,2})/\sqrt{\sigma^2_{1/p_T,1}+\sigma^2_{1/p_T,2}}$.
Results are shown for four geometries: data before alignment (black dotted lines), 
data with combined method alignment (red solid), combined method MC (green dashed), 
and ideal MC (blue dash-dotted).
}
\label{fig:split}
\end{center}
\end{figure}
The track segments were reconstructed independently.
Figure~\ref{fig:split} shows the difference between upper and lower
portions of tracks for all five track parameters. 
There is significant improvement due to tracker alignment, with good
agreement between data and MC simulations.
The results of the combined method are approaching those of a MC 
simulation with ideal detector geometry.
The $p_T$ measurements are most sensitive 
to the strip part of the tracker, while the other four parameters are 
dominated by the alignment of the pixel detector.  
The normalized distributions in Fig.~\ref{fig:split} also show that
the error estimates on the track parameters are in good agreement
with predictions from MC simulations.
Figure~\ref{fig:split-vspt} shows the dependence of the resolution on 
$p_T$ for the track parameters $d_{xy}$ and $1/p_T$.

\clearpage

\subsection{Systematic misalignment studies}
\label{sec:systematics}

A global translation and rotation of the whole tracker is an example
of a trivial transformation which leaves the $\chi^2$ value of
Eq.~(\ref{eq:chisq}) unchanged.
This transformation has no effect on the internal alignment and is easily
resolved by a suitable convention when defining the global reference frame.
The convention that the center-of-gravity of all modules 
coincides with the design position was used in this work. 
A similar convention was used for the rotation of the tracker.

There are, however, non-trivial transformations,
so-called ``weak'' modes, that also preserve the $\chi^2$ value of Eq.~(\ref{eq:chisq}) 
and are of larger concern.
The presence of weak modes in the geometry resulting from
track-based alignment was investigated following the approach
described in Ref.~\cite{babaralignment}. Nine systematic distortions, in
$\Delta r$, $\Delta\phi$, and $\Delta z$ as a function of $r$, $\phi$, 
and $z$ were applied to the aligned geometry. Studies were performed
with both global and local alignment methods and the results presented here
were obtained with the global method. 
The systematically misaligned geometries were used as a starting point and the 
alignment procedure described in Section~\ref{sec:stat} was repeated.
The nine geometries obtained after the alignments were then
compared to the original aligned geometry to see 
if the distortions can be recovered by the alignment procedure.

\begin{figure}[htb]
\begin{center}
\includegraphics*[width=0.85\linewidth]{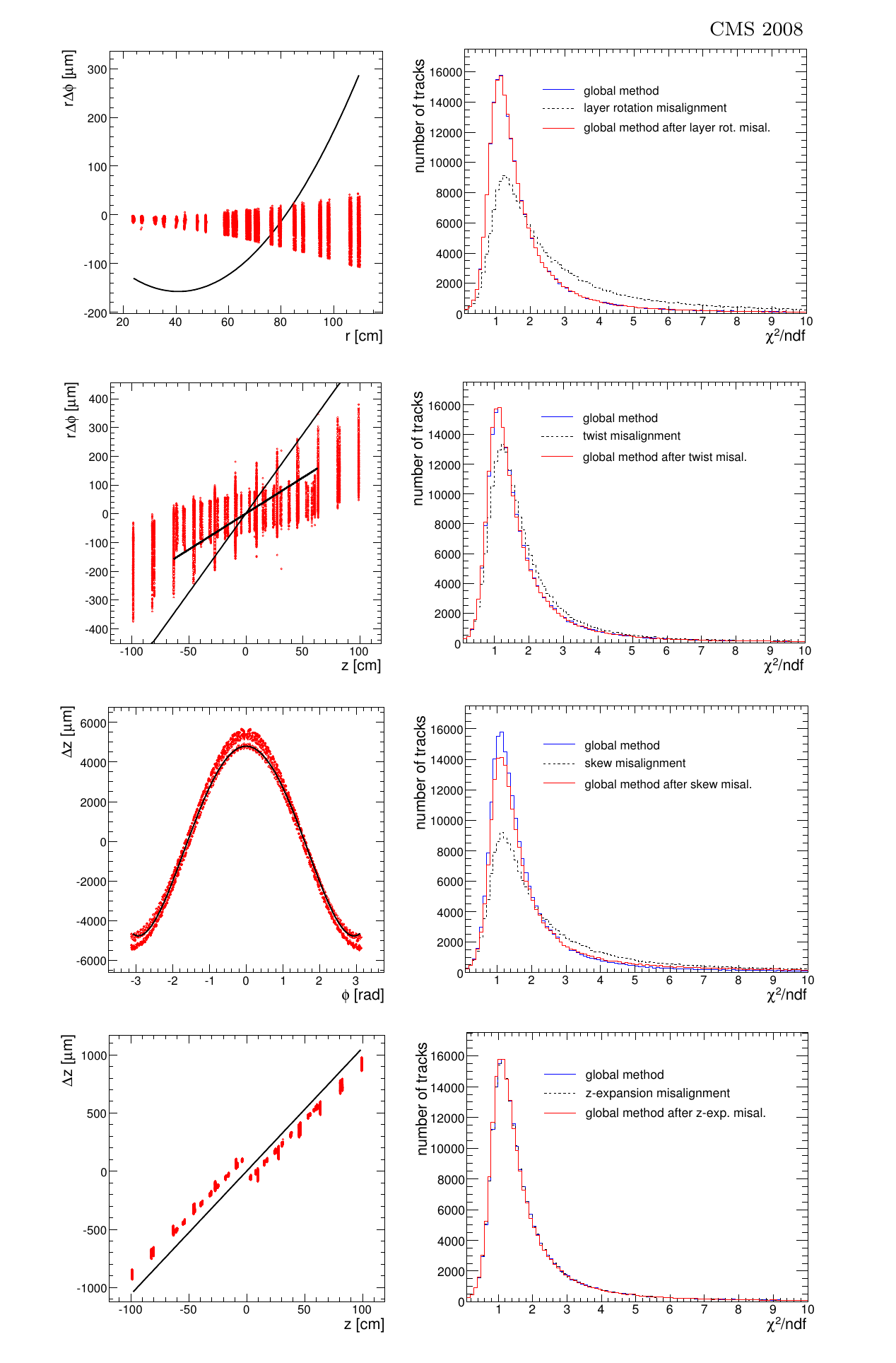}
\caption{
Comparison of the position of the TIB and TOB modules with respect to the geometry 
obtained with the global method after applying systematic distortions (black solid lines) 
and after alignment (red dots) is shown on the left for layer rotation (top row), 
twist (second row), skew (third row), and $z$-expansion (bottom row) weak modes.
For the twist misalignment, the shorter and longer solid lines correspond to the modules in
the inner layer of the TIB and in the outer layer of the TOB, respectively.
The plots on the right show the distributions of the corresponding track $\chi^2/\mathrm{ndf}$ 
after the alignment with the global method (blue solid line), after introducing the
systematic misalignment (black dashed line), and after re-aligning 
(red solid line, below the blue solid line).
}
\label{fig:syst_global}
\end{center}
\end{figure}

The results were analyzed separately for the pixel (BPIX and FPIX),
barrel strip (TIB and TOB), and forward strip (TID and TEC) sub-detectors.
The remaining displacements of the modules in the TIB and TOB, which are
expected to have the best illumination from cosmic ray tracks, are shown 
in Fig.~\ref{fig:syst_global} for four systematic deformations: 
the layer rotation ($\Delta \phi=c_1+c_2 r$), the twist deformation
($\Delta \phi = c_1 z$), the skew ($\Delta z=c_1\cos\phi$), and the $z$-expansion ($\Delta z = c_1 z$).
The layer rotation can be well recovered in alignment with cosmic ray tracks,
while the scatter in the module position difference is an evidence for weak
modes in other projections. 
The twist and the skew deformations are reflected to a lesser degree in track $\chi^2$, 
and therefore are only partially recovered.
Finally, as the distribution of the track $\chi^2$ changes only marginally with respect 
to the $z$-expansion, this deformation cannot be recovered using cosmic ray tracks only.

\clearpage

Radial deformations are generally recovered using constraints from overlapping modules.
Also, mid-plane differences between the upper and lower legs of cosmic ray tracks are sensitive
to deformations that differently affect the upper and lower parts of the tracker,  
including the layer rotation and telescope ($\Delta z = c_1 r$) modes.
However, more subtle deformations, including those discussed above, 
may be difficult to recover with cosmic ray tracks alone. Additional information
with tracks from LHC beam interactions, which should be uniform in $\phi$, and
with additional constraints from the verticies and the masses of resonances, 
may provide better sensitivity to those systematic deformations. 
In principle, the Laser Aligment System (Section~\ref{sec:las}) and hierarchical
survey measurements provide complementary information,
but evaluating systematic biases in those measurements 
is a challenging task.

\subsection{Study of detector geometry}
\label{sec:geometry}

Two sets of comparisons between geometries obtained from track-based alignment 
with data have been performed: a comparison between the geometries from the local 
and the global methods, and a comparison between the geometry obtained with the 
combined method with respect to the design one.
A comparison between two geometries is done after correcting for an 
overall residual shift and rotation of the whole detector, 
or a sub-detector, with respect to its center-of-gravity.

\begin{figure}[tbh]
\begin{center}
\includegraphics*[width=0.4\linewidth]{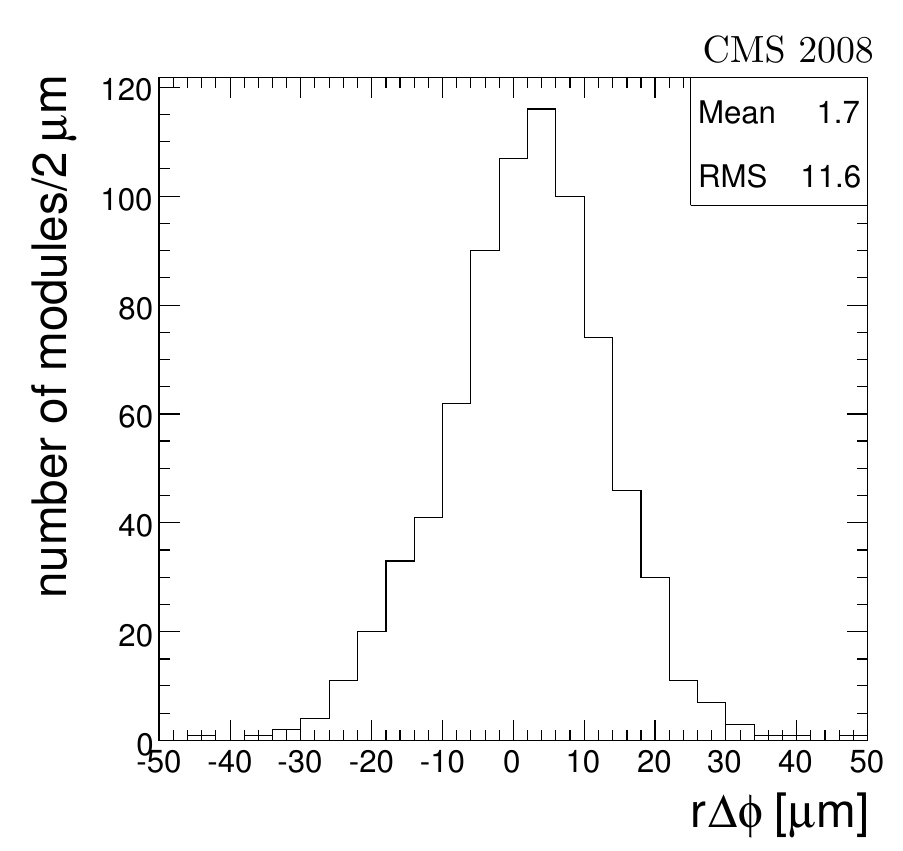}
\caption{
Differences of the BPIX module positions $r(\phi_{\rm local}-\phi_{\rm global})$
as obtained with the local and global methods.
}
\label{fig:rdphi_PXB_localglobal}
\end{center}
\end{figure}

Figure~\ref{fig:rdphi_PXB_localglobal} shows the distribution
of the differences between the $r\phi$ positions
of the 768 BPIX modules obtained with the local and the global
methods. The RMS of the distribution is about 12 $\mu$m.
The distribution of the differences between the $z$ positions is similar once 
the two main weak modes, $\Delta z$ vs.~$z$ and $\Delta z$ vs.~$\phi$, 
are properly identified and removed by the means of the fit 
with the functions shown in the left of Fig.~\ref{fig:syst_global}.
Similar module-by-module comparisons were performed on larger structures,
like the TIB or TOB, where weak modes cannot be easily removed, and
on coordinates for which the track-based alignment with cosmic rays has a limited
sensitivity. This led to distributions characterized by a larger spread. 
This is also confirmed by alignment on simulated data and limits the
accuracy of the absolute position determination. 

\begin{figure}[tbh]
\begin{center}
\includegraphics*[width=\linewidth]{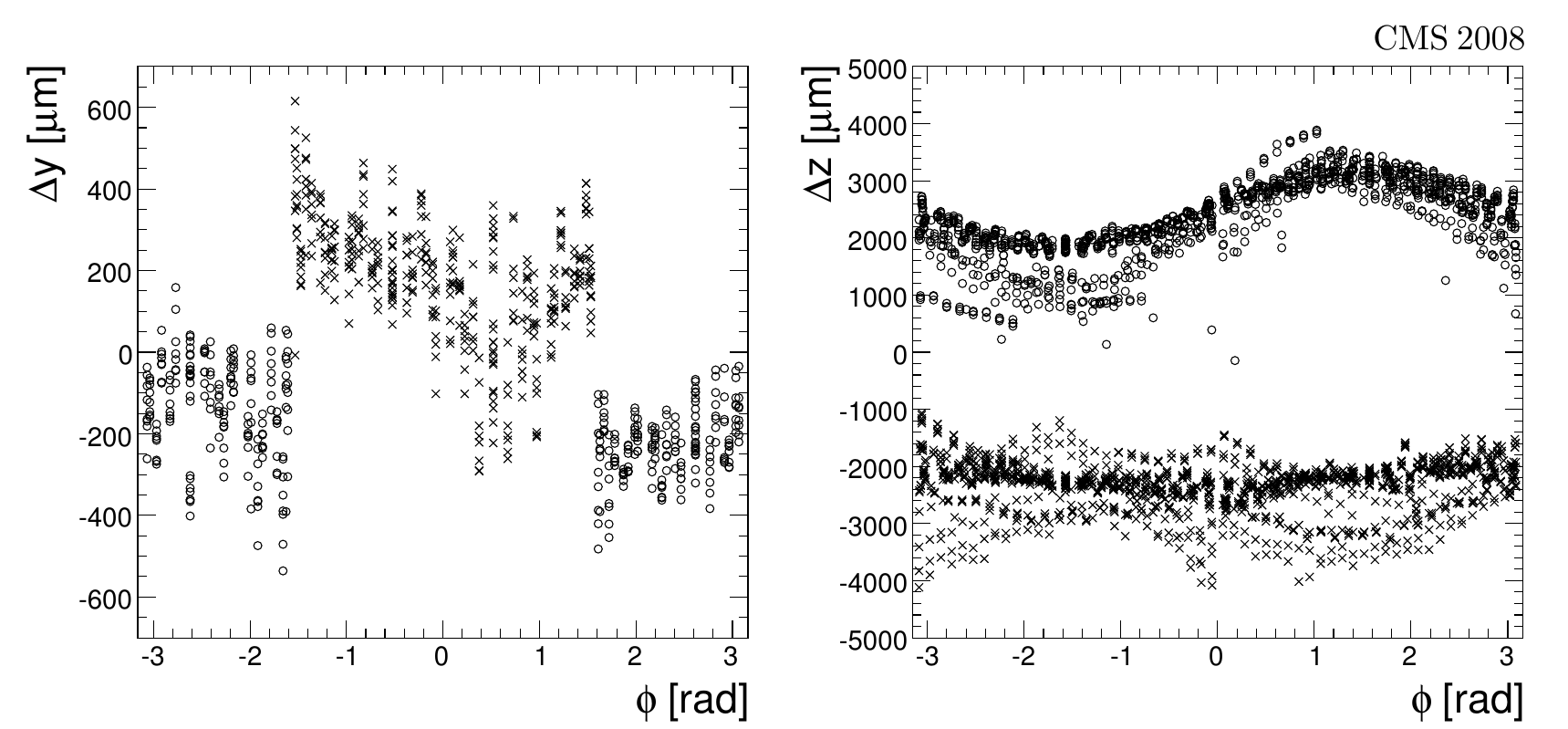}
\caption{ Comparison of the position of the modules in the combined
  method geometry with respect to the design one:
  $(y_{design}-y_{comb})$ for the BPIX modules (left) and
  $(z_{design}-z_{comb})$ for the TIB modules (right) as a function of
  $\phi$.  In the plot on the left crosses (circles) represent modules of the
  positive (negative) $x$ BPIX half-barrel.  In the plot on the right crosses
  (circles) represent modules of the positive (negative) $z$ TIB
  half-barrel.  }
\label{fig:combined_vs_design}
\end{center}
\end{figure}

Assuming that the geometry of the tracker is best described by
the one obtained with the combined method, a comparison with the design
geometry indicates that the two BPIX half-barrels are shifted along the
vertical axis by about 0.4 mm and the two half-barrels of the TIB have 
an extra separation along the $z$ axis of about 5 mm. 
Both displacements are mechanically possible. 
The large displacement of the TIB half-barrels is supported by the optical
survey measurements described in Ref.~\cite{tifalignment}.
Figure~\ref{fig:combined_vs_design} shows the displacement in the
$y$- and in the $z$-coordinates of the BPIX and TIB modules with
respect to the design position.
In the case of the TIB, a large scatter of the $z$-coordinate of the
modules with respect to their design position is observed. 
The modulation of the displacement as a function of the $\phi$
position of the module can be explained by the presence of 
the weak mode skew in the combined geometry, as discussed
in the previous section.


\subsection{Validation with the Laser Alignment System}
\label{sec:las}

An independent test of the silicon tracker alignment was provided by
the Laser Alignment System (LAS), which uses a system of 40 infrared
laser beams ($\lambda=1075$\,nm) to survey the position of the
large-scale structure elements of the tracker. The LAS measurements
are available for 434 silicon strip modules, which are distributed over
eight azimuthal sectors. The light is detected directly on the
active area of the silicon sensors and therefore provides excellent
beam position resolution with respect to the module sensitive
area. For each TEC disk, there are 16 modules distributed uniformly in
$\phi$ at two radial positions that are intersected by laser beams.
Furthermore, the LAS is capable of measuring the relative orientation
of both endcaps and the TIB and TOB half-barrels.  The TID and the
pixel detector are not included in the~LAS.

Operation of the LAS during data-taking was mainly devoted to 
commissioning. In total, 31 out of
the 40 laser beams were fully operational during this period. The available
data were used to calculate the relative positions of the TEC disks. 
Using LAS measurements in the alignment procedure by means 
of additional terms to Eq.~(\ref{eq:chisq}) is not discussed in this paper.

For the calculation of the alignment parameters from LAS data, the
TEC disks were treated as solid objects, each with one rotational and two
translational degrees of freedom. Additional degrees of freedom resulted
from the fact that the beam directions may deviate from their nominal
values at the level of 1~mrad. The calculation was based on an
analytical approach for solving the $\chi^2$-minimization problem for
the laser hit residuals. 

\begin{figure}[htb]
\begin{center}
\includegraphics*[height=5.3cm]{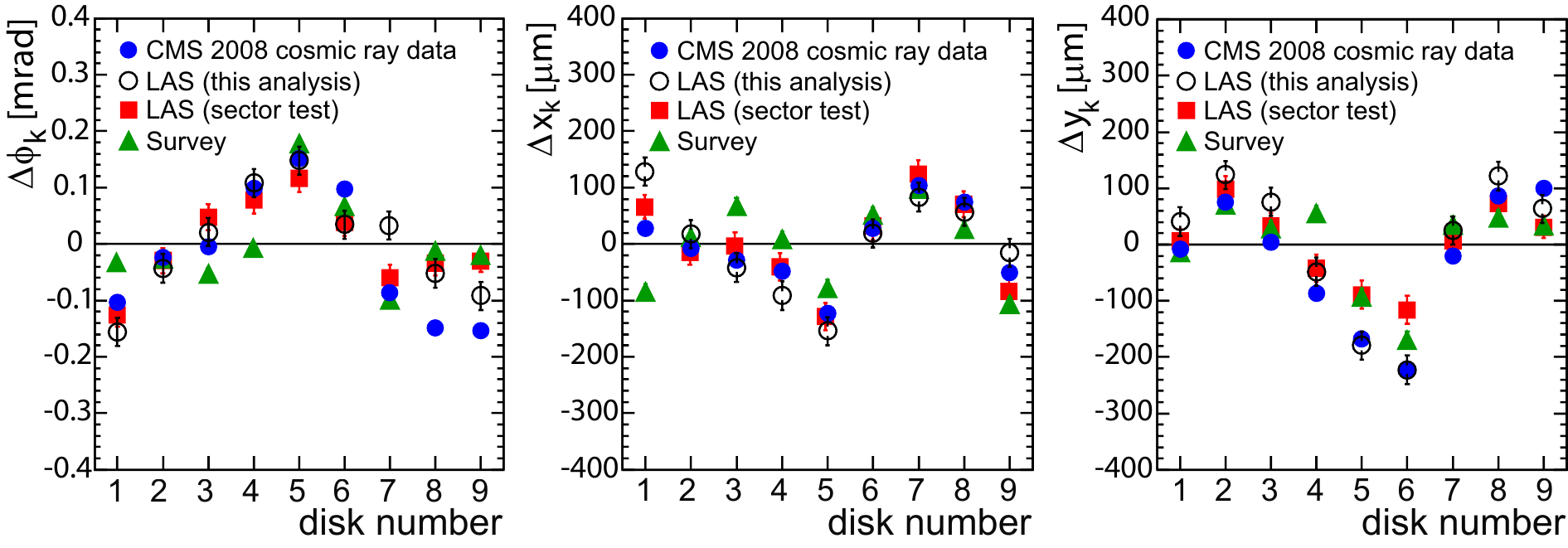}
\caption{
The three alignment parameters
for positive $z$ TEC disks measured 
using cosmic ray data (solid blue circles),
by the LAS in this analysis (open black circles),
during the LAS sector test~\cite{tifalignment} (red squares),
and by optical survey (green triangles): 
rotation around global $z$ (left) and  translation
in global $x$ (middle) and in global $y$ (right).
}
\label{fig:lasTec}
\end{center}
\end{figure}

For comparison, the position of the TEC disks was also calculated from the
results of the track-based alignment using the combined method, where
only the position of the modules illuminated by the LAS beams were
considered, for consistency. Figure~\ref{fig:lasTec} compares the
results of the measurements for positive $z$ side TEC disks from the
cosmic ray data analysis, the LAS data analysis, the LAS sector
test~\cite{tifalignment}, and optical survey. All data are in 
good agreement. However, it is difficult to estimate the systematic
uncertainties in the results for the track-based alignment due to the effects
discussed in Section~\ref{sec:systematics}. Small movements of
particular disks are expected between the sector test and optical
survey measurements and the CRAFT results from the LAS
and the track-based alignment.
The optical surveys were done with the endcaps in a vertical orientation 
whereas they have a horizontal orientation within the tracker.

\begin{figure}[htb]
\begin{center}
\includegraphics*[height=5.3cm]{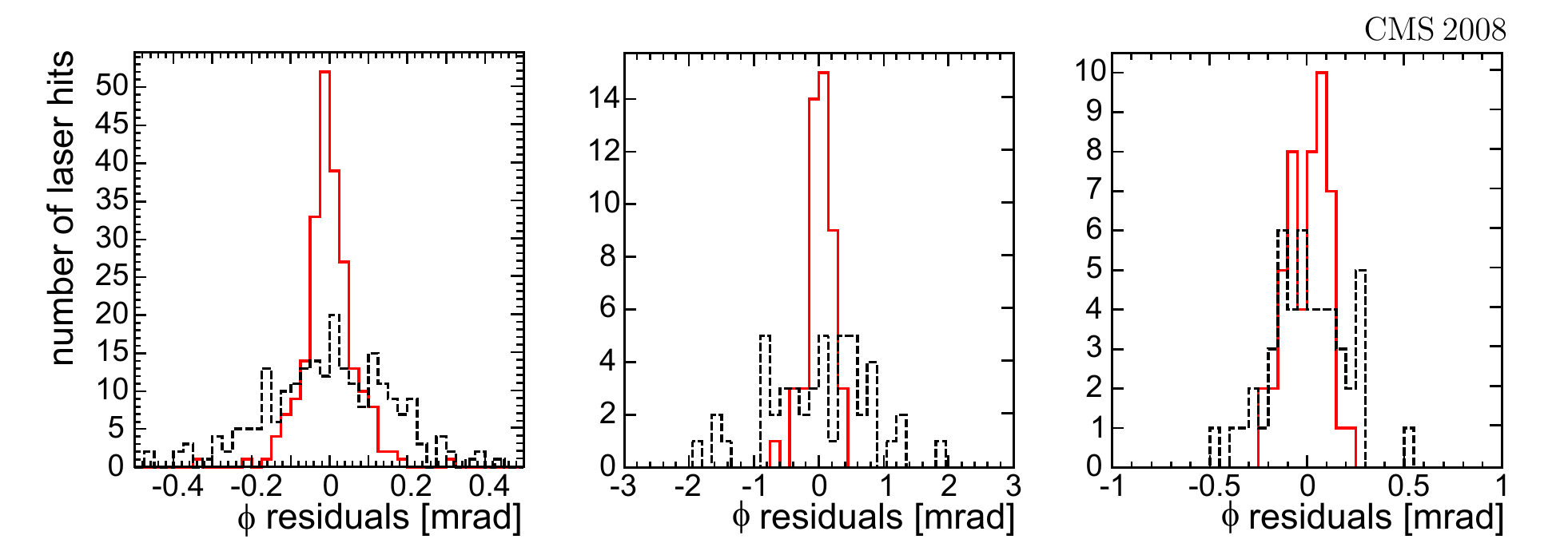}
\caption{
Laser hit residuals ($\phi_\mathrm{nominal} -\phi_\mathrm{measured}$) 
for the TEC (left), TIB (middle), and TOB (right), using the geometry before alignment
(dashed black line) and the geometry obtained from the combined
alignment method (solid red line).}
\label{fig:lasResiduals}
\end{center}
\end{figure}

As a comparison to track-based results, the laser hit residuals, defined as
the difference between the nominal and the measured laser hit position along
the $\phi$-coordinate, have been determined from the tracker 
geometry before alignment as well as assuming the geometry from the combined method.
Corrections have been made for deviations of the beams from their nominal
orientations. The results are shown in Fig.~\ref{fig:lasResiduals}. Assuming
the aligned geometry as measured with tracks, the overall RMS of the residual
distribution decreases significantly with respect to the before alignment case
(cf.~Table~\ref{tab:las}). For the modules in the TEC, the RMS decreases by
more than a factor of two, while the effect in the TOB is slightly
smaller. The largest relative improvement is observed for the TIB
modules. However, the resulting RMS of 200~$\mu$rad for TIB modules remains
larger than in the other sub-detectors. The observed residual widths may 
reflect an insufficient calibration of the lasers and a non-optimized laser
beam correction procedure. Furthermore, weak modes for which the track-based
alignment with cosmic rays has limited sensitivity may contribute to the
residual distributions observed in the LAS studies.

\begin{table*}[tbh]
\caption{
RMS of laser hit residuals ($\phi_\mathrm{nominal} -\phi_\mathrm{measured}$)
for the TEC, TIB, and TOB before and after alignment with the combined alignment
method.
}
\label{tab:las}
\begin{center}
\begin{tabular}{|l|c|c|c|}
\hline
                          &  TIB      & TOB           & TEC \\
                          &  [$\mu$rad]   &  [$\mu$rad]   & [$\mu$rad] \\
\hline                   
before align.  &  790          & 200           & 160 \\
aligned        &  200          & 110           & 70 \\
\hline
\end{tabular}
\end{center}
\end{table*}

\section{Summary and Discussion}
\label{sec:discuss}

In summary, results of the first full alignment of the CMS tracker have been presented. 
Two algorithms have been used to determine the positions of all 16\,588 silicon modules. 
The two alignment methods have been combined sequentially to take into account 
most effectively both global and local correlations of module positions.
The results are based on the analysis of  about three million cosmic ray tracks 
recorded with a 3.8\,T magnetic field.

The precision of the detector positions with respect to particle trajectories
has been derived from the distribution of the median of the cosmic muon 
track residuals to be on average 3--4\,$\mum$ RMS in the barrel and 3--14\,$\mum$ RMS in the 
endcaps in the most sensitive coordinate. These results are supported by the difference 
of the residuals in regions where modules overlap within a layer, thus reducing effects 
of multiple scattering in the  calculation of the residuals. 
Nevertheless, additional small systematic effects may still be present.
The achieved tracker resolution in all five track parameters has been checked
with a study of the two halves of a cosmic ray track
and compared with predictions from a detailed detector simulation. 
The measured resolutions are close to those that would be observed 
in a detector with perfectly placed modules.

Overall, a significant improvement in track fit performance 
compared either to using the before alignment or survey geometry has been observed. 
The resulting aligned geometry allows a study of the assembly precision 
of the individual sub-detectors. A comparison of module positions obtained 
with complementary methods supports findings based on track residuals. 
Nonetheless, certain deformations in the geometry  that do not change 
the $\chi^2$ of the tracks, cannot be ruled out due to the largely vertical 
nature of the cosmic track data.

Clear improvement in the LAS residuals is observed when using the
result of the track based alignment. 
The stand-alone LAS data analysis agrees well with 
that from previous tests and with survey alignment.
The operation of the LAS has shown that the laser beams operate properly. 

Experience gained in the alignment analysis of the silicon
modules with cosmic ray particles is valuable in preparation for the
CMS tracker alignment with the data from LHC collisions, which 
is critical to achieving the physics goals of the CMS detector.
Integration of measurements from cosmic ray and collision tracks, 
LAS, and survey will be critical for the optimal tracker alignment
with the first data expected from LHC beam collisions.
Because the track parameter resolutions are close to the design performance  
with cosmic ray data already,  statistical uncertainties in track parameters 
are not expected to be considerably affected by alignment at CMS startup 
with data from beam collisions.

\section{Acknowledgments}
\label{sec:acknowledgment}
We thank the technical and administrative staff at CERN and other CMS
Institutes, and acknowledge support from: FMSR (Austria); FNRS and FWO
(Belgium); CNPq, CAPES, FAPERJ, and FAPESP (Brazil); MES (Bulgaria);
CERN; CAS, MoST, and NSFC (China); COLCIENCIAS (Colombia); MSES
(Croatia); RPF (Cyprus); Academy of Sciences and NICPB (Estonia);
Academy of Finland, ME, and HIP (Finland); CEA and CNRS/IN2P3
(France); BMBF, DFG, and HGF (Germany); GSRT (Greece); OTKA and NKTH
(Hungary); DAE and DST (India); IPM (Iran); SFI (Ireland); INFN
(Italy); NRF (Korea); LAS (Lithuania); CINVESTAV, CONACYT, SEP, and
UASLP-FAI (Mexico); PAEC (Pakistan); SCSR (Poland); FCT (Portugal);
JINR (Armenia, Belarus, Georgia, Ukraine, Uzbekistan); MST and MAE
(Russia); MSTDS (Serbia); MICINN and CPAN (Spain); Swiss Funding
Agencies (Switzerland); NSC (Taipei); TUBITAK and TAEK (Turkey); STFC
(United Kingdom); DOE and NSF (USA). Individuals have received support
from the Marie-Curie IEF program (European Union); the Leventis
Foundation; the A. P. Sloan Foundation; and the Alexander von Humboldt
Foundation.


\bibliography{auto_generated}
\cleardoublepage\appendix\section{The CMS Collaboration \label{app:collab}}\begin{sloppypar}\hyphenpenalty=500\textbf{Yerevan Physics Institute,  Yerevan,  Armenia}\\*[0pt]
S.~Chatrchyan, V.~Khachatryan, A.M.~Sirunyan
\vskip\cmsinstskip
\textbf{Institut f\"{u}r Hochenergiephysik der OeAW,  Wien,  Austria}\\*[0pt]
W.~Adam, B.~Arnold, H.~Bergauer, T.~Bergauer, M.~Dragicevic, M.~Eichberger, J.~Er\"{o}
% ...
\vskip\cmsinstskip
% ...

\end{sloppypar}
\end{document}